\PassOptionsToPackage{pdfpagelabels=false}{hyperref} 
\documentclass[a4paper,fleqn,usenatbib]{mnras}
\usepackage{txfonts}
\usepackage{mathptmx}
\usepackage{txfonts}
\usepackage[T1]{fontenc}
\usepackage{ae,aecompl}

\usepackage{graphicx}	
\usepackage{amssymb}	

\title[The multi-line slope method]{The multi-line slope method for the measure of the effective magnetic field of
the cool stars: an application to the
solar like cycle of $\epsilon$\,Eri.}
\author[C. Scalia et al.]{
C.~Scalia,$^{1,2}$\thanks{E-mail: cesare.scalia@oact.inaf.it}
F.~Leone,$^{1,2}$
M.~Gangi,$^{1,2}$
M.~Giarrusso,$^{1,2}$
M.J.~Stift$^{3}$
\\
$^{1}$Universit\`a di Catania, Dipartimento di Fisica e Astronomia, Sezione Astrofisica,
Via S. Sofia 78, I--95123 Catania, Italy \\
$^{2}$INAF - Osservatorio Astrofisico di Catania, Via S. Sofia 78, I--95123 Catania, Italy\\
$^{3}$Armagh Observatory, College Hill, Armagh BT61 9DG, Northern Ireland, UK\\
}

\date{Accepted 2017 August 10. Received 2017 August 10; in original form 2017 May 16}

\pubyear{2017}

\begin{document}
\label{firstpage}
\pagerange{\pageref{firstpage}--\pageref{lastpage}}
\maketitle

\begin{abstract}
A method for the determination of integrated longitudinal stellar fields from low-resolution spectra
is the so-called \textit{slope method}, which is based on the regression of the Stokes\,$V$ signal
against the first derivative of Stokes\,$I$. Here we investigate the possibility to extend this technique to
measure the magnetic fields of cool stars from high resolution spectra.
For this purpose we developed a multi-line modification to the slope method, called \textit{multi-line slope method}.
We tested this technique by analysing synthetic spectra computed with the {\sc  Cossam} code and real observations
obtained with the high resolution spectropolarimeters Narval, HARPSpol and \textit{Catania Astrophysical
Observatory Spectropolarimeter} (CAOS). We show that the multi-line slope method is a fast alternative to the
Least Squares Deconvolution (LSD) technique for the measurement of the effective magnetic fields of cool stars.
Using a Fourier transform on the effective magnetic field variations of the star $\epsilon$\,Eri, we find
that the long term periodicity of the field corresponds to the 2.95\,yr period of the stellar dynamo, revealed
by the variation of the activity index.
\end{abstract}




\section{Introduction}

Magnetic fields are among the most elusive physical phenomena that play an important
role in the physics of the atmosphere of late-type stars; their direct observation is
difficult since their effects are usually hidden in the typical noise of astronomical
observations.

Indicators of stellar magnetism are chromospheric Ca lines \citep{Schrijver1989} and 
coronal X-ray emission \citep{Pevtsov2003} both of which however are not directly 
related to the field strength. As reported by \citet{Judge2012}, an empirical
correlation between Zeeman signals of magnetic field and chromospheric indices has 
been found in the Sun and in other stars, like $\xi$ Boo \citep{Morgenthaler2010}.

Measuring and monitoring the behaviour of the magnetic field of late type stars is 
important in order to better understand dynamo theories. It is commonly accepted 
that physical processes at the origin of the magnetic field in cool stars are the 
same as in the Sun, but with a different set of parameters, such as temperature, 
gravity and stellar rotation \citep{Reiners2012}. The magnetic field is an essential
ingredient to chromospheric and coronal heating, it also plays a role in the 
accretion of circumstellar material onto the stellar surface \citep{Bouvier2007}, 
in the theory of the formation of exoplanets, and in star-planet interaction 
\citep{Preusse2006,Strugarek2015}. The impact of magnetic fields on stellar activity can mimic the modulation of the stellar radial 
velocity caused by the presence of exoplanets \citep{Dumusque2012}, leading to false detections
\citep{Queloz2001}, among them the planetary systems of HD\,219542 
\citep{Desidera2004}, HD\,200466 \citep{Carolo2014} and HD\,99492 \citep{Kane2016}. 

The polarisation signal due to the Zeeman effect is so small in cool stars that the
current instrumentation is not able to detect it in individual spectral lines. For this
reason several techniques are being developed in order to detect magnetic signals. 
\citet{Semel1996} proposed a multi-line technique to add the polarisation signal
originating from several spectral lines into one pseudo profile, with an higher signal to
noise. The most used method to add spectral profiles is the \textit{Least square
deconvolution} \citep{Donati1997}. However, there are other techniques such
the \textit{Principal Component Analysis} \citep{Semel2009} or the \textit{Zeeman
Component Decomposition} \citep{Sennhauser2010}.

In this work we present an alternative method to measure the integrated longitudinal 
magnetic field strength of cool stars. We extend to high resolution spectroscopy the 
method applied by \citet{Bagnulo2002} to low resolution spectra. Our technique, 
hereafter called \textit{multi-line slope method}, allows to measure the field from 
the slope of Stokes\,$V$ versus the spectral derivative of Stokes\,$I$. We apply the 
multi-line slope method to high resolution data of the K2V star $\epsilon$\,Eri. We 
analyse data from archives of NARVAL \citep{aur2003} and HARPSpol \citep{Snik2011, 
Piskunov2011} and new observations obtained with the spectropolarimeter CAOS 
\citep{Leone2016}.

The paper develops as follow. In Sect.\,\ref{observation} we describe the data-set 
of observations of $\epsilon$\,Eri. In Sect.\,\ref{secteor} we describe the general 
slope method, in Sect.\,\ref{multislope} we introduce the multi-line approach and 
we subsequently test it numerically. Sect.\,\ref{compareLSD} presents a comparison
of the multi-line slope method with the LSD technique. In Sect.\,\ref{star} we 
discuss the measurements of the magnetic field of $\epsilon$\,Eri and in 
Sect.\,\ref{Conclusion} we report the final conclusions of the work.

\section{Observations}
\label{observation}

\subsection{CAOS}
We started to observe the star $\epsilon$\,Eri using the spectropolarimeter 
\textit{Catania Astrophysical Observatory Spectropolarimeter} (CAOS) in 2014. The instrument is fiber linked to the 
0.91\,m telescope of the Catania Astrophysical Observatory (\textit{G. M. Fracastoro} Stellar Station, Serra La Nave, Mt. Etna,
 Italy).

Stokes\,$V$ observations are recorded through a Savart plate and a $\lambda/4$ wave-plate,
with exposures at angles of 45\degr and 135\degr . Following \citet{Tinbergen1992}, single
exposures of a polarimeter are affected by two functions $G$ and $F(t)$, time independent
and time dependent respectively:

\begin{equation}
{\rm i_o=0.5\,(I+V)\,G_o\,F(t) \,\,\,\,; \,\,\,\,
i_s=0.5\,(I-V)\,G_s\,F(t) }
\end{equation}

From the ratios of the single exposures:
 
\begin{equation}\label{pol1}
{\rm R_V^{4}=\frac{i_{1o}}{i_{2o}}\frac{i_{2s}}{i_{1s}}\frac{i_{4o}}{i_{3o}}\frac{i_{3s}}{i_{4s}}\,\,;\,\, R_{N}^{4}=\frac{i_{1o}}{i_{2o}}\frac{i_{2s}}{i_{1s}}\frac{i_{3o}}{i_{4o}}\frac{i_{4s}}{i_{3s}} }
\end{equation}
We compute Stokes\,$V$ and the null profile \citep{Donati1997}:

\begin{equation}\label{stokes}
{\rm \frac{V}{I}=\frac{R_V-1}{R_V+1}\,\,;\,\,\frac{N}{I}=\frac{R_N-1}{R_N+1}}
\end{equation}

The stability of the wavelength calibration in time is of crucial importance for the 
combination of the exposures. We calibrate a Thorium-Argon reference lamp to a 
precision of $10^{-4}${\AA}. In order to achieve high accuracy of the measurements, 
we take particular care of the thermal stability of the spectrograph which is better 
than 0.01\,K.

\subsection{Data from archives}

In this work we use all the available spectropolarimetric observations of
$\epsilon$\,Eri. The archival data consist of HARPSpol and NARVAL observations. HARPSpol in located on the 3.6 telescope
 in La Silla while Narval is located on the 2.2\,m Telescope Bernard Lyot.
 
The raw science and the calibration files of the HARPSpol 
observations were downloaded from the ESO 
archive\footnote{\url{http://archive.eso.org/eso/eso_archive_main.html}} and they refer to observations
 taken in January 2010 and February 2011. 
We performed the data reduction by using IRAF packages and we computed 
Stokes\,$V$ through Eq.\,\ref{pol1} and Eq.\,\ref{stokes}.

NARVAL data were downloaded from the PolarBase
database\footnote{\url{http://polarbase.irap.omp.eu/}} \citep{Petit2014} and they refer to six different epochs:
 January 2007, January 2008, January 2010, October 2011, October 2012 and October 2013.

The complete logbook of the observations is in Table\,\ref{table:magneticmeasure}.

\section{The slope method}
\label{secteor}

The slope method is based on the assumption of weak field approximation. If we 
assume that the Zeeman pattern can be approximated by a classical triplet and 
if the Zeeman separation $\Delta \lambda_B$ is small compared to the intrinsic 
broadening of a spectral line, the emergent circular polarisation from a point 
on the surface of a star can be written as function of the spectral derivative of Stokes\,$I$ \citep{Unno1956}:

\begin{equation}\label{eq:stelsuf}
{\rm V(\lambda,\theta)=\Delta \lambda_B\, cos\phi\,\frac{dI(\lambda,\theta)}{d\lambda} }
\end{equation}
where $\phi$ and $\theta$ are respectively the angle between the magnetic field 
vector and the line of sight and the angle between the local surface normal 
and the line of sight and the Zeeman separation is given by:
\begin{equation}
{\rm \Delta \lambda_B=4.67 \, 10^{-13} \, g_{{\rm eff}} \, \lambda^2 \, B }
\end{equation}
where ${\rm g_{eff}}$ is the effective Land\'e factor, $B$ is the field strength 
expressed in Gauss and $\lambda$ is the wavelength in {\AA}ngstrom.

Previous Eq.\,\ref{eq:stelsuf} is strictly valid for a single element of the stellar surface. \citet{Landstreet1982}
noted that the extension to the stellar (spatially unresolved)
case runs into difficulties related to 1) stellar rotation which Doppler shifts the local profiles, and 2)
because both $\Delta \lambda_B$ and the angle $\phi$ vary over the
visible stellar disk. Assuming that the velocity broadening is small compared to the
intrinsic (magnetic) broadening, he showed that the observed Stokes\,V is related
to the observed Stokes\,$I$ by:
\begin{equation}\label{eqfond}
{\rm \frac{V}{I}=-4.67 \, 10^{-13} \, g_{{\rm eff}} \, \lambda^2 B_{eff} \, \frac{dI}{d\lambda} \, \frac{1}{I}}
\end{equation}
where $\rm B_{eff}$ is the integral over the visible hemisphere of the magnetic field component
along the line of sight, usually called the \textit{effective magnetic field}, expressed in Gauss.

\citet{Bagnulo2002} introduced the idea to measure the effective magnetic field of faint stars through the application of
Eq.\,\ref{eqfond} to low resolution (R$\le$5000) spectra. However, this equation holds if line profiles are shaped by the magnetic field and not by
the instrumental broadening. A condition that let the method suggested by Bagnulo  and co-workers well suited for  early-type stars, whose Balmer lines dominate
at low resolution. For example, \citet{Kolenberg2009} found no evidence of magnetic fields in RR-Lyr stars, \citet{Leone2007} observed some magnetic A-type stars to test the capability of the William Hershel Telescope
 of measuring the stellar magnetic field, \citet{Leone2011} gave an upper limit for the magnetic field in central stars of planetary
nebulae and \citet{Hubrig2016} detected magnetic fields in Wolf-Rayet stars.

\section{The multi-line slope method}\label{multislope}

Eq.\,\ref{eqfond}  holds for any spectral line shaped by the magnetic field
and, in principle, the simultaneous application to a large number of spectral lines can result in a very
sensitive measurement of the field, as it is usual for radial velocities when thousands of spectral lines result in a  cross-correlation function.

\citet{Bagnulo2002} have pointed out that Eq.\,\ref{eqfond} is restricted to unblended lines, so that the method
cannot be applied to metal lines as observed at low resolution spectroscopy.
We propose an extension of the slope method to the unblended lines of late-type stars observed at high resolution.
This multi-line approach presents the advantage of taking into account the correct  line-by-line $g_{\rm eff}$ value.
\cite{Leone2007} have adopted an average values, while \citet{Bagnulo2012} pointed out how the use of an average Land{\'e} factor
limits the precision of the measurement of the effective magnetic field from circular polarisation, since the value of
$g_{\rm eff}$ varies among the lines; for many lines, the actual circular polarisation will vary
from the average by up to 25\%.

The process of selection started with a synthetic spectrum computed
by SYNTHE \citep{Kurucz1993, Sbordone2004}. First we removed all the lines weaker than the
noise level and the lines in the region of strong lines, like H$\alpha$ and H$\beta$, and
telluric lines. The exclusion of very broad and strong spectral lines is mainly justified by the presence of line cores dominated by saturation
and not by magnetic fields.

In order to select the most sensitive transitions, we included only lines 
whose effective Land\'e factor is larger than 0.7. Atomic parameters are taken from the
VALD database \citep{Piskunov1995}.

\begin{figure}
\centering
\includegraphics[trim=1.0cm 0.5cm 0.5cm
10.5cm,clip=true,width=8cm]{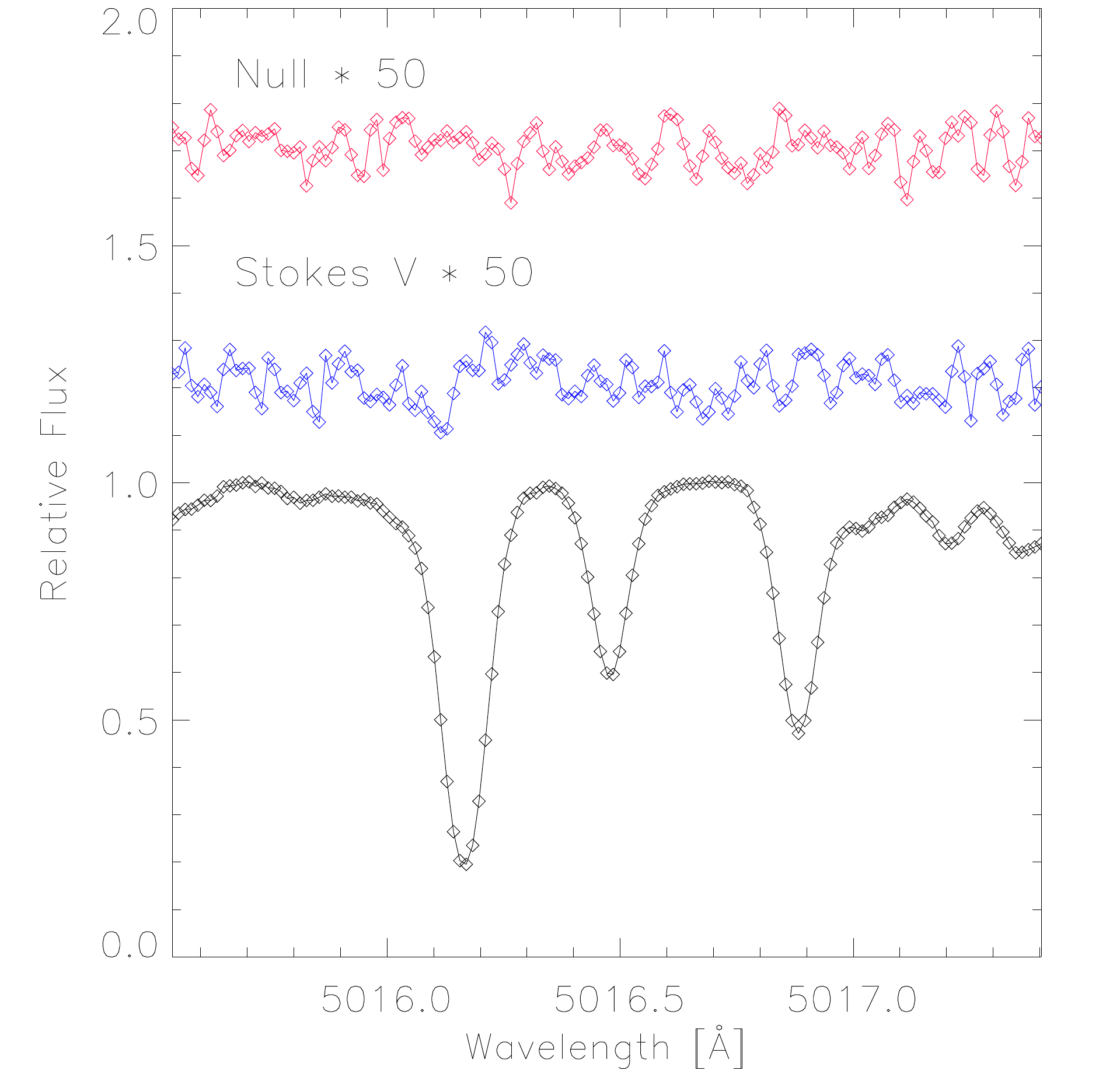}
\includegraphics[trim=1.0cm 0.5cm 0.5cm
10.5cm,clip=true,width=8cm]{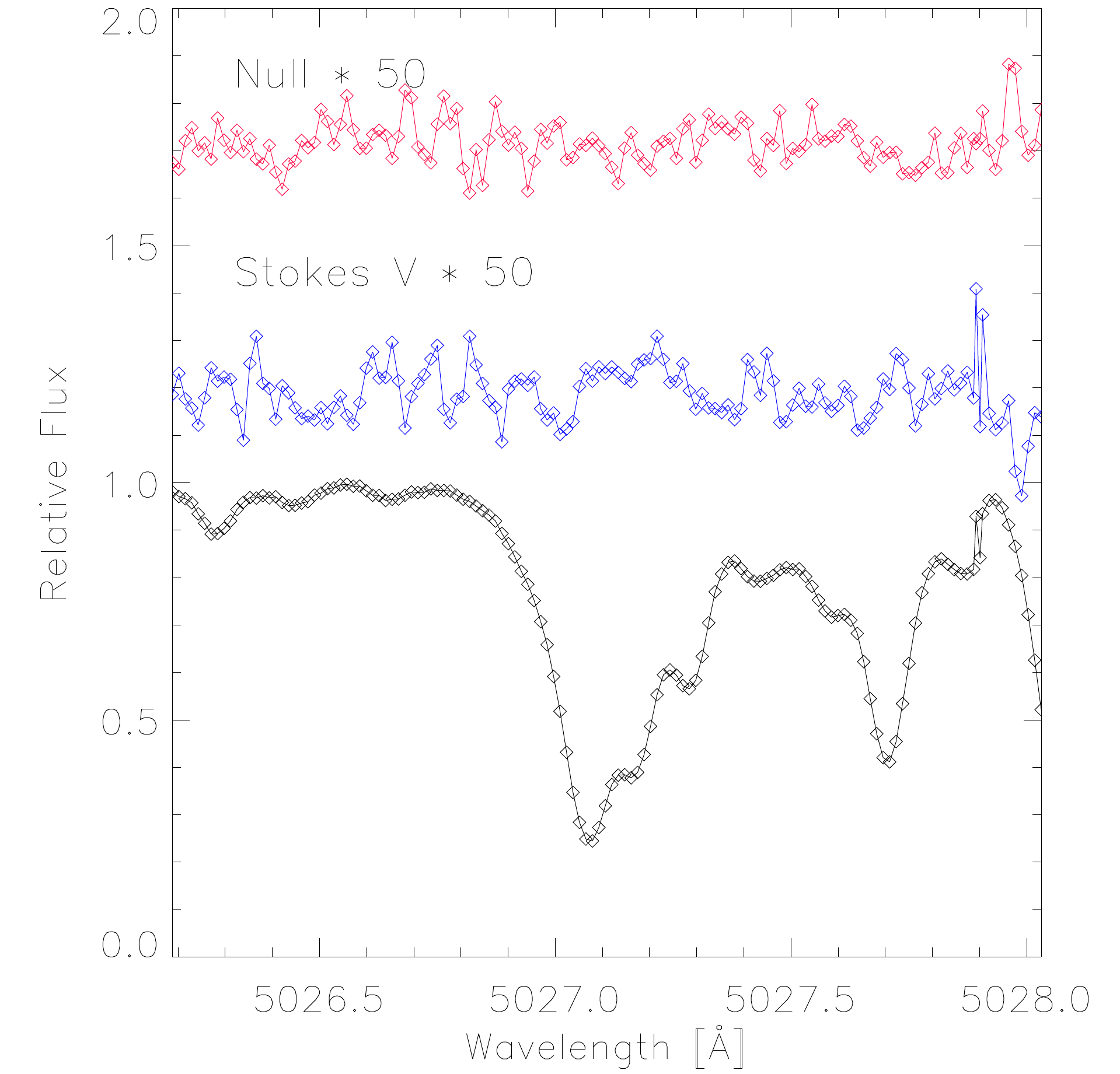}
\includegraphics[trim=1.0cm 0.5cm 0.5cm
10.5cm,clip=true,width=8cm]{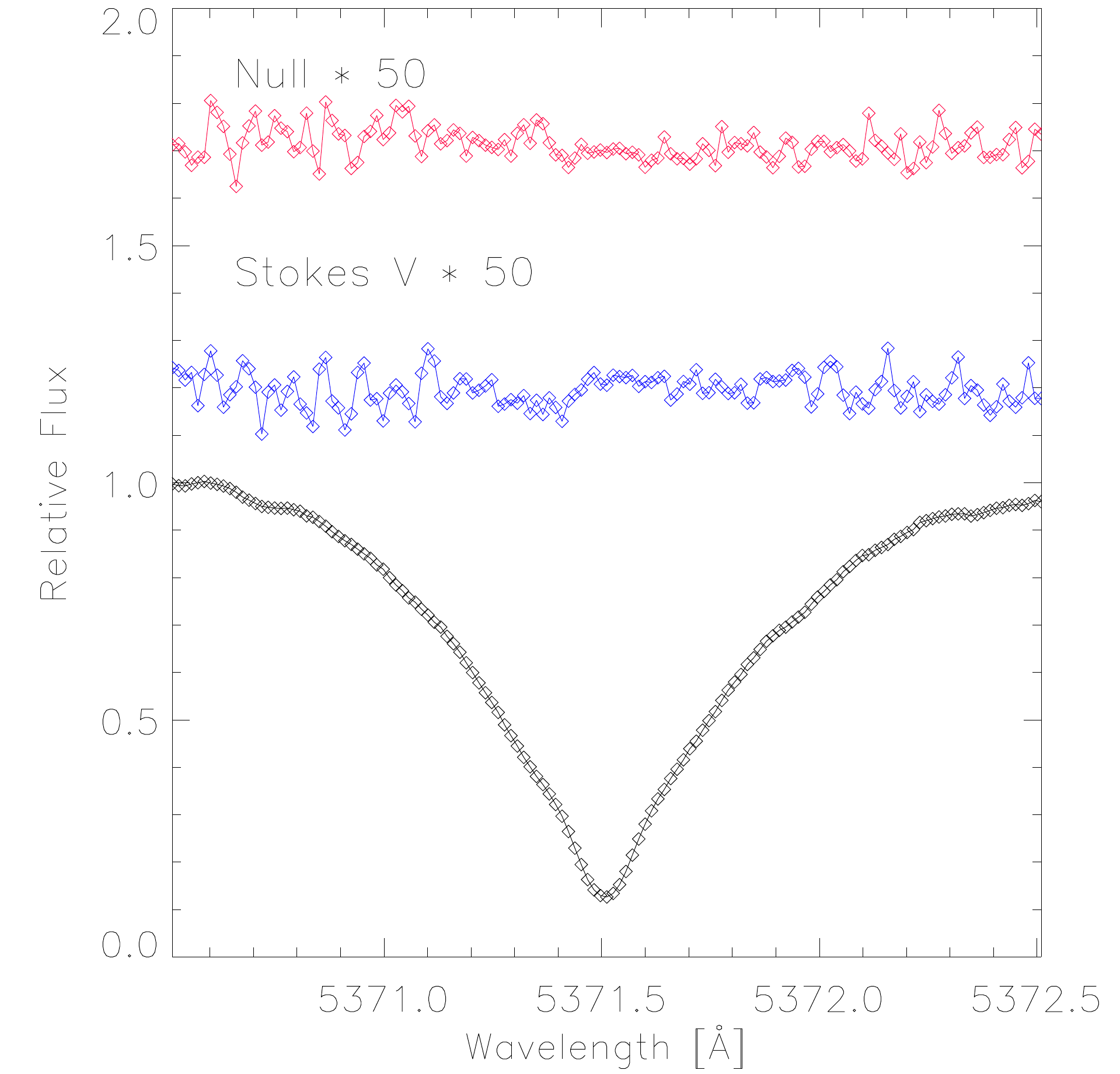}

\caption{Example of selections on the spectra of $\epsilon$\,Eri observed with HARPSpol on 5 January 2010. From the top to the bottom we reported examples of selected unblended lines, unselected blend lines and unselected strong lines.}
\label{fig:lineexample}
\end{figure}

In order to find and remove blends, we evaluated the FWHM (in ${\rm km}$ ${\rm s^{-1}}$) and the
position of the centroid of each line through a gaussian fit and we discarded all 
the transitions whose radial velocity or the FWHM was distant more than 3$\sigma$ 
from the averaged value. This is possible since data were acquired with an \'echelle 
spectrograph in which $\frac{\Delta \lambda}{\lambda} \approx constant$.
Fig.\,\ref{fig:lineexample} shows examples of selected and unselected spectral lines.
 In average a thousand of spectral lines are selected for the measure. 

\begin{figure}
\centering
\includegraphics[trim=1.0cm 0.5cm 0.5cm
10.5cm,clip=true,width=8cm]{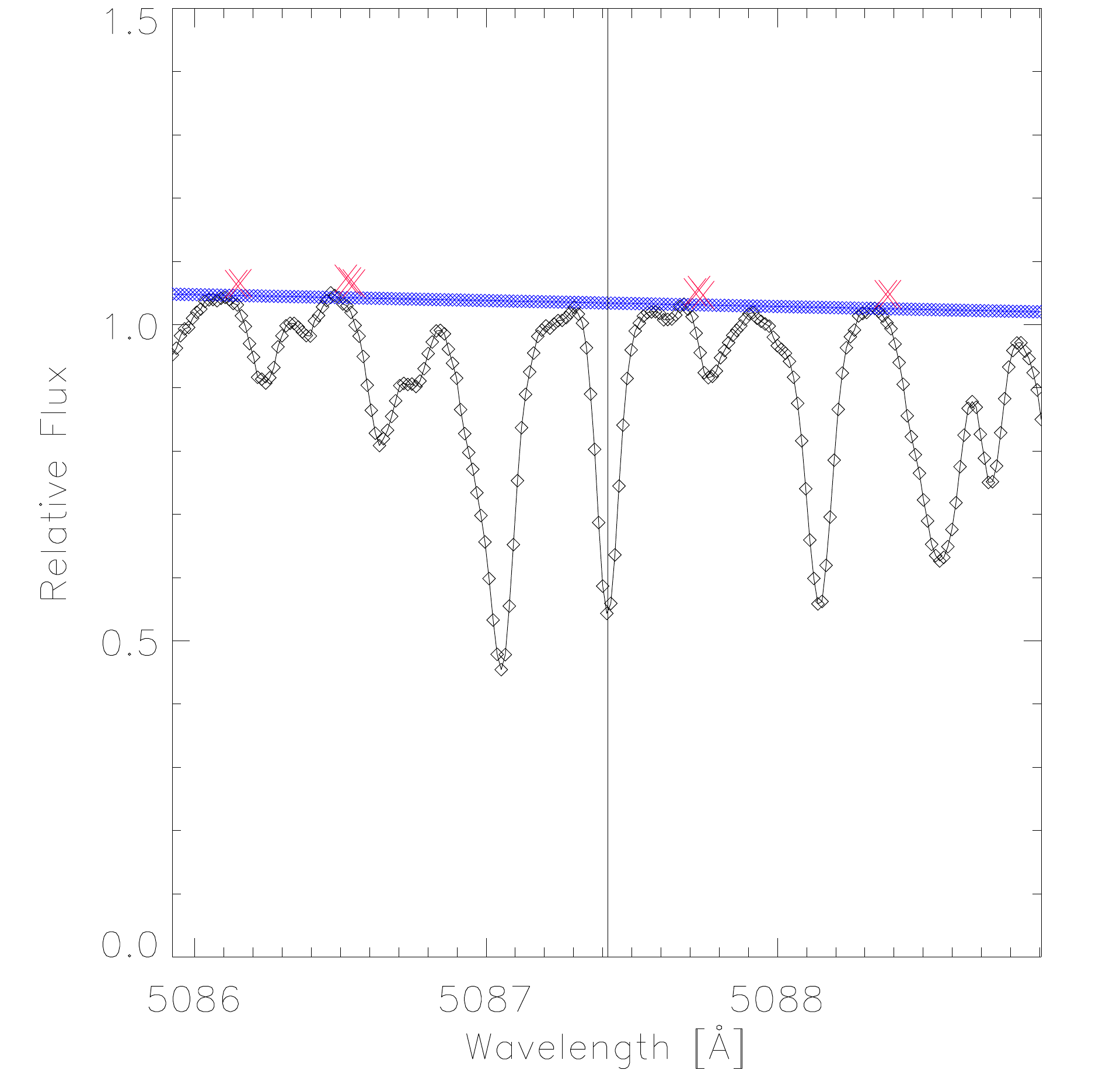}
\caption{Example of normalisation of Stokes\,$I$; the linear pseudo continuum is reported in blue and the points used for the fit are in red.}
\label{fig:esenorm}
\end{figure}

Another possible source of error is the normalisation to the
continuum \citep{Bagnulo2012}. In order to limit the impact, we computed the pseudo continuum
level of each line through the linear fit of the highest ten points in a region of 3\,{\AA} centered on the wavelength of the
transition, half on the right and half on the left (Fig. \ref{fig:esenorm}).

For each line (with index $j$) we computed the quantity
\begin{equation}\label{xslope}
{\rm x_{ij}={\rm -4.67\,10^{-13}\,\lambda_{0j}^{2}\,
g_{eff_{j}}\frac{1}{I_{ij}(\lambda)}\frac{dI_{ij}(\lambda)}{d\lambda_{ij}}} }
\end{equation}
where $i$ extends over the pixels. It is possible to note that Eq.\,\ref{xslope} allows the use of the effective Land\'e 
factor $g_{{\rm eff}}$ of each particular line instead of the average value. The 
spectral derivative ${\rm \frac{dI_{ij}(\lambda)}{d\lambda_{ij}}}$ was computed using a 
3-point Lagrange interpolation. Spikes in the spectra, due for example to cosmic 
rays, can affect the measurement of the magnetic field. To avoid this we performed 
a clipping of the null profile, rejecting all $V/I$ points whose $N/I$ is more 
than $5\sigma$ from the average \citep{Bagnulo2006}.

We determined the magnetic field through minimisation of $\chi^2$ as given by
\begin{equation}\label{eqfin}
{\rm \chi^{2}=\sum_{ij}\frac{\left(y_{ij}- B_{eff}\,\, x_{ij}-b\right)^{2}}{\sigma_{ij}^{2}}}
\end{equation}
where $b$ is a constant term related to the residual instrumental polarisation 
\citep{Bagnulo2002}. We used $y_{ij}=\frac{V_{ij}}{I_{ij}}$ for the measurement of the 
magnetic field and $y_{ij}=\frac{N_{ij}}{I_{ij}}$ for the estimation of systematic 
errors. This is possible because the null profile is, in principle, related to 
systematic errors due to observations or to the data reduction procedure.
Finally, the total error was given by the quadratic sum of the systematic error 
and the standard error of the fit. In the case of a good measurement, we expected 
that the slope of $\frac{N_{ij}}{I_{ij}}$ is zero within the error range 
\citep{Leone2007}.

\begin{figure}
\centering
\includegraphics[trim=0.cm 1.3cm 1.5cm 9.7cm,clip=true,width=7.1cm]{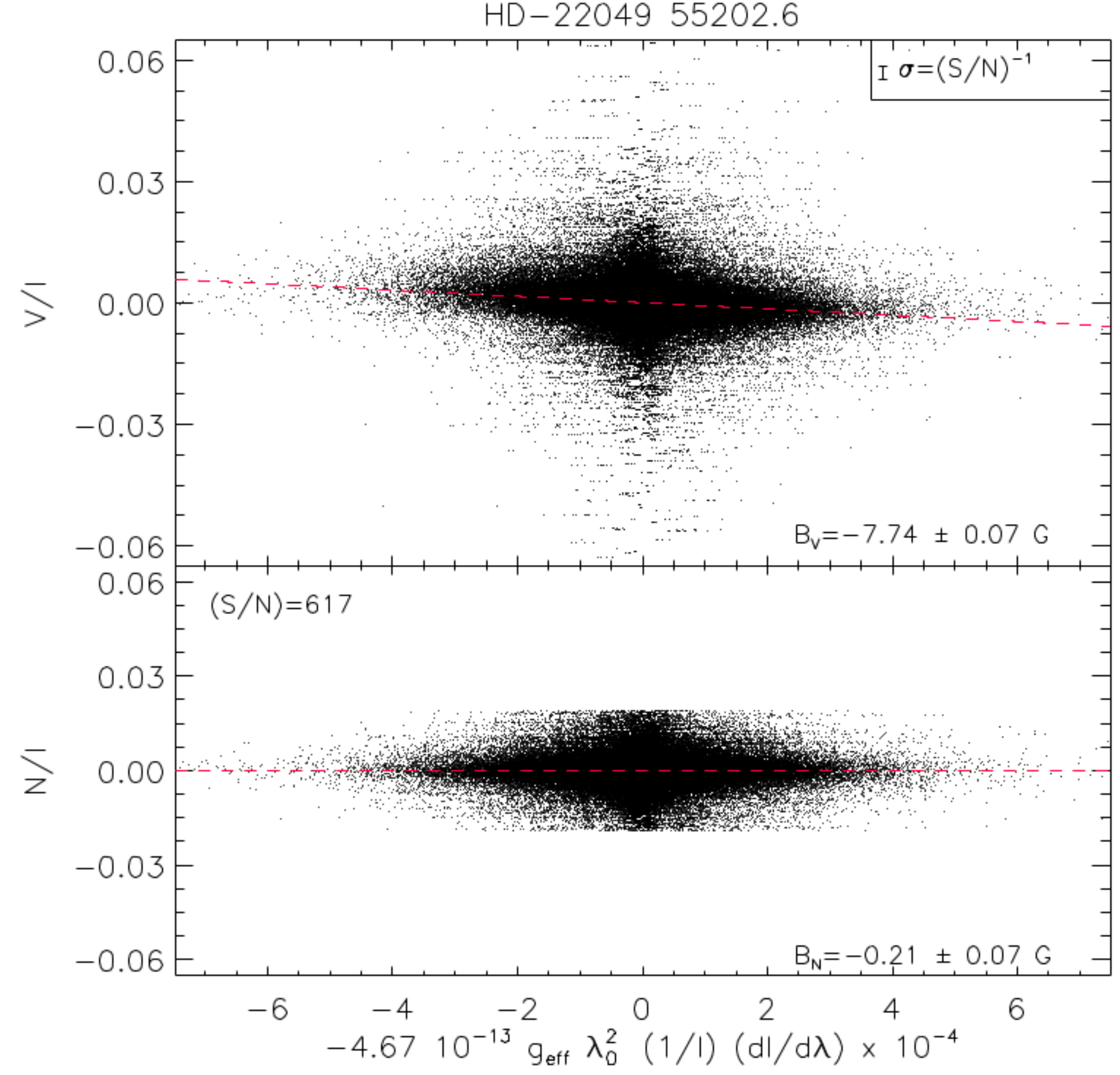}
\includegraphics[trim=0.cm 1.3cm 1.5cm 9.7cm,clip=true,width=7.1cm]{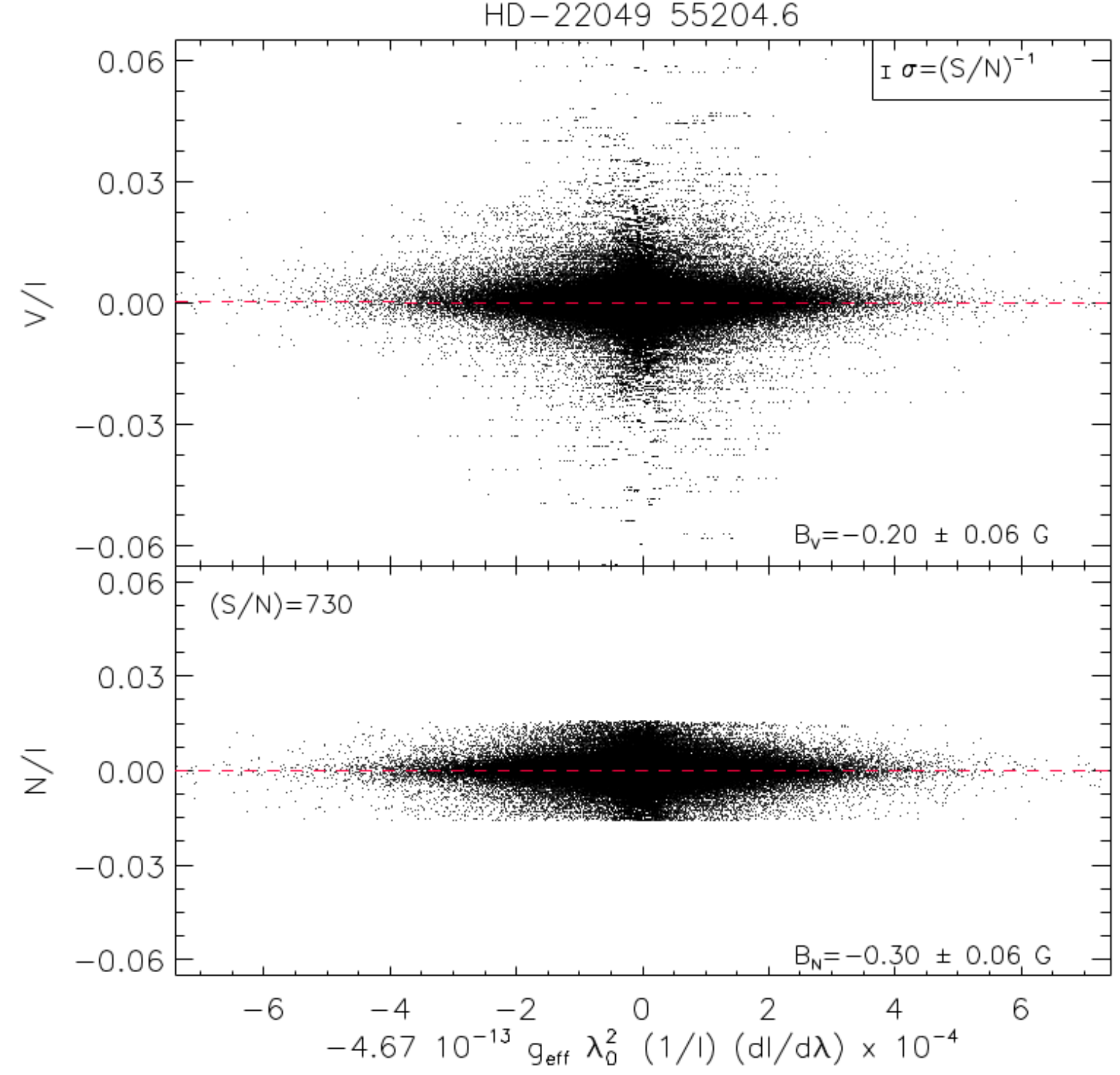}
\includegraphics[trim=0.cm 1.3cm 1.5cm 9.7cm,clip=true,width=7.1cm]{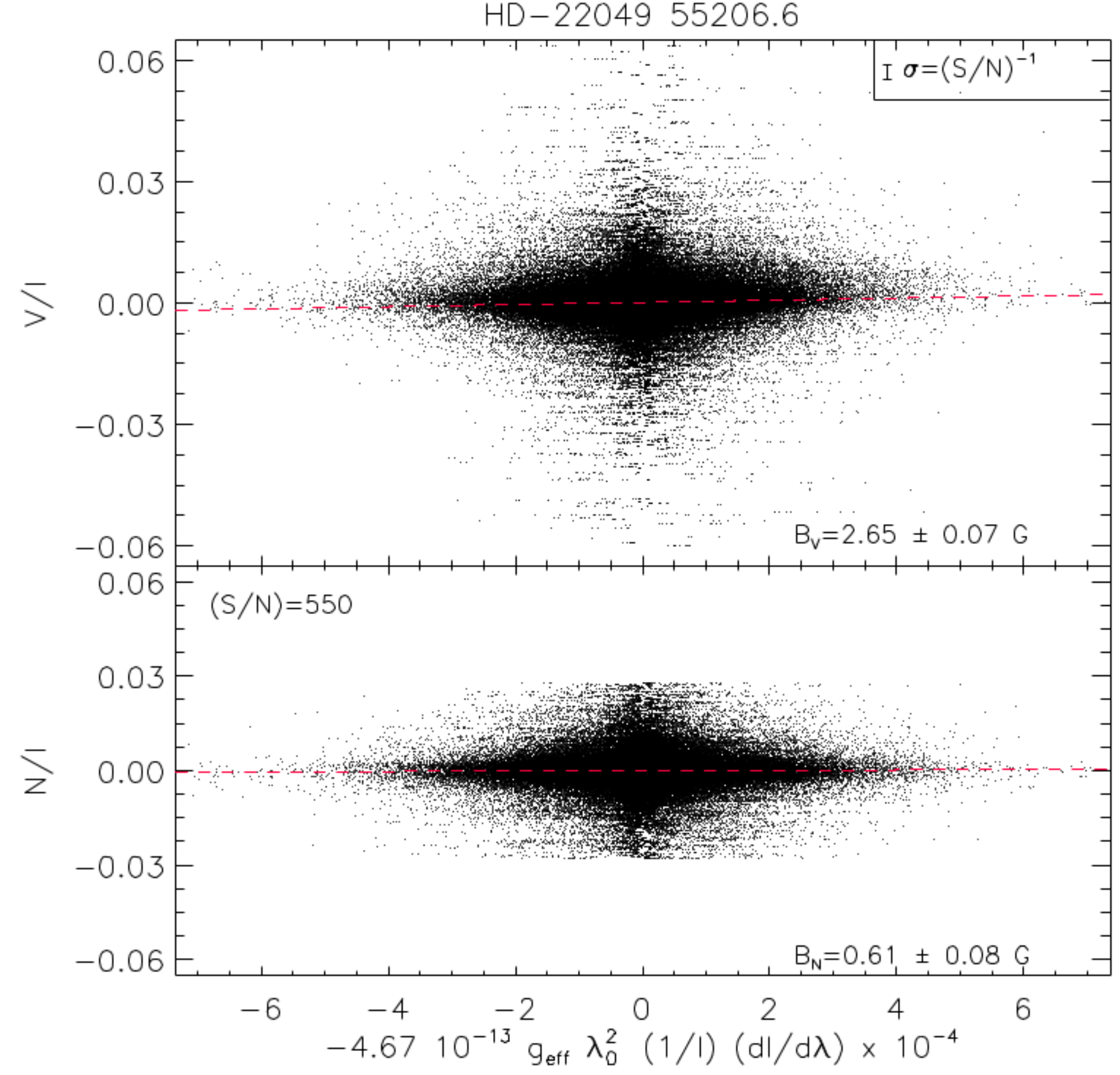}
\caption{Magnetic field measurements of $\epsilon$\,Eri from HARPSpol data. In the 
top panels we plot the Stokes\,$V$ profiles and in the bottom panels the null profiles,
both as a function of the spectral derivative of Stokes\,$I$. The presence of the 
magnetic field is reflected in the slope of the distribution. The flat distribution 
of the null profile confirms the good quality of the measure.
Figure refers, from top to the bottom, to the observation made on the nights of 5, 7 and 9 January 2010. The sizes of the typical
 error bar are showed in the figures; S/N is calculated from the standard deviation of the Null profile's points.}

\label{fig:example_slope}
\end{figure}

Fig.\,\ref{fig:example_slope} shows an example of the magnetic field measurements obtained with the application of the multi-line slope method to high resolution spectra of the cool 
star $\epsilon$\,Eri observed by HARPSpol using a total of 3900 lines.

\subsection*{Numerical tests}

 \begin{table*}
    \caption{Results of effective magnetic field measurement from the multi-line slope method ($B_{{\rm ms}}$) and from the slope method and from the slope method applied on all the simulated data points in the spectral region ($B_{\rm {slope}}$) vs rotational velocity. The input effective magnetic field is $B_{\rm inp} = -6.45$ ${\rm G}$.}
    \label{table:vel}
    \centering
	\resizebox{\textwidth}{!}{

   \begin{tabular}{ c@{   }r@{   }r@{   }r r@{   }r@{   }r@{   }r@{   }r r@{   }r@{   }r@{   }r@{   }r r@{   }r@{   }r@{   }r@{   }r  }
        \hline\hline\\[-2.0ex]
       \multicolumn{4}{c}{ } &
       \multicolumn{5}{c}{CAOS}  &
       \multicolumn{5}{c}{NARVAL} &
     \multicolumn{5}{c}{HARPSpol}\\

      \multicolumn{4}{c}{} &
       \multicolumn{5}{c}{R=$55000$} &
      \multicolumn{5}{c}{R=$65000$} &
      \multicolumn{5}{c}{R=$115000$} \\

    \multicolumn{4}{c}{} &
     \multicolumn{5}{c}{2.5 px per FWHM} &
      \multicolumn{5}{c}{2.5 px per FWHM} &
      \multicolumn{5}{c}{4.1 px per FWHM} \\

    \multicolumn{4}{c}{} &
     \multicolumn{5}{c}{} &
     \multicolumn{5}{c}{} &
    \multicolumn{5}{c}{} \\

     \multicolumn{4}{c}{} &
  \multicolumn{5}{c}{} &
  \multicolumn{5}{c}{} &
    \multicolumn{5}{c}{} \\
    \hline\\[-1.0ex]
${\rm v}$ ${\rm sin(i)}$    & &&& B$_{\rm ms}$ &$\Delta$B$_{\rm inp - ms} $ & Lines & B$_{\rm slope}$                    & $\Delta$B$_{\rm inp - slope} $     & B$_{\rm ms}$  & $\Delta$B$_{\rm inp - ms} $   &Lines                & B$_{\rm slope}$            & $\Delta$B$_{\rm inp - slope} $     & B$_{\rm ms}$     & $\Delta$B$_{\rm inp - ms} $        & Lines         & B$_{\rm slope}$             & $\Delta$B$_{\rm inp - slope} $   \\
$\rm [km $ ${\rm s^{-1}]}$  & &&& ${\rm [G]}$  & \multicolumn{1}{c}{(\%)}   &       & \multicolumn{1}{c}{${\rm [G]}$}     & \multicolumn{1}{c}{(\%)}         & ${\rm [G]}$   & \multicolumn{1}{c}{(\%)}    &                 & \multicolumn{1}{c}{${\rm [G]}$}  & \multicolumn{1}{c}{(\%)}         & ${\rm [G]}$      & \multicolumn{1}{c}{(\%)}         &              & \multicolumn{1}{c}{${\rm [G]}$}       & \multicolumn{1}{c}{(\%)}\\
       \hline\\[-1.0ex]
0  & &&& -6.59 & \multicolumn{1}{c}{2}  & \multicolumn{1}{c}{697} & -8.60 & \multicolumn{1}{c}{33} & -6.69 & \multicolumn{1}{c}{4} & \multicolumn{1}{c}{697} & -8.46 & \multicolumn{1}{c}{31} & -6.33 & \multicolumn{1}{c}{-2} & \multicolumn{1}{c}{697} & -7.72 & \multicolumn{1}{c}{20} \\
3  & &&& -7.28 & \multicolumn{1}{c}{13} & \multicolumn{1}{c}{688} & -9.18 & \multicolumn{1}{c}{42} & -7.21 & \multicolumn{1}{c}{12} & \multicolumn{1}{c}{688} & -9.12 & \multicolumn{1}{c}{41} & -7.08 & \multicolumn{1}{c}{10} & \multicolumn{1}{c}{688} & -8.83 & \multicolumn{1}{c}{37} \\
6  & &&& -8.13 & \multicolumn{1}{c}{26} & \multicolumn{1}{c}{615} & -10.04 & \multicolumn{1}{c}{56} & -8.24 & \multicolumn{1}{c}{28} & \multicolumn{1}{c}{615} & -10.15 & \multicolumn{1}{c}{57} & -8.20 & \multicolumn{1}{c}{27} & \multicolumn{1}{c}{615} & -10.17 & \multicolumn{1}{c}{58} \\
9  & &&& -8.49 & \multicolumn{1}{c}{32} & \multicolumn{1}{c}{537} & -10.57 & \multicolumn{1}{c}{64} & -8.66 & \multicolumn{1}{c}{34} & \multicolumn{1}{c}{537} & -10.67 & \multicolumn{1}{c}{65} & -8.57 & \multicolumn{1}{c}{33} & \multicolumn{1}{c}{537} & -10.53 & \multicolumn{1}{c}{63} \\
12 & &&& -8.53 & \multicolumn{1}{c}{32} & \multicolumn{1}{c}{411} & -10.72 & \multicolumn{1}{c}{66} & -8.51 & \multicolumn{1}{c}{32} & \multicolumn{1}{c}{411} & -10.70 & \multicolumn{1}{c}{66} & -8.44 & \multicolumn{1}{c}{31} & \multicolumn{1}{c}{411} & -10.48 & \multicolumn{1}{c}{63} \\
15 & &&& -8.74 & \multicolumn{1}{c}{35} & \multicolumn{1}{c}{301} & -10.74 & \multicolumn{1}{c}{66} & -8.69 & \multicolumn{1}{c}{35} & \multicolumn{1}{c}{301} & -10.67 & \multicolumn{1}{c}{65} & -8.50 & \multicolumn{1}{c}{32} & \multicolumn{1}{c}{301} & -10.43 & \multicolumn{1}{c}{62} \\
18 & &&& -8.85 & \multicolumn{1}{c}{37} & \multicolumn{1}{c}{236} & -10.80 & \multicolumn{1}{c}{67} & -8.64 & \multicolumn{1}{c}{34} & \multicolumn{1}{c}{236} & -10.57 & \multicolumn{1}{c}{64} & -8.54 & \multicolumn{1}{c}{32} & \multicolumn{1}{c}{236} & -10.43 & \multicolumn{1}{c}{62} \\
21 & &&& -8.99 & \multicolumn{1}{c}{39} & \multicolumn{1}{c}{187} & -10.74 & \multicolumn{1}{c}{67} & -9.08 & \multicolumn{1}{c}{41} & \multicolumn{1}{c}{187} & -10.66 & \multicolumn{1}{c}{65} & -8.80 & \multicolumn{1}{c}{36} & \multicolumn{1}{c}{187} & -10.39 & \multicolumn{1}{c}{61} \\
24 & &&& -9.43 & \multicolumn{1}{c}{46} & \multicolumn{1}{c}{146} & -10.77 & \multicolumn{1}{c}{67} & -9.31 & \multicolumn{1}{c}{44} & \multicolumn{1}{c}{146} & -10.57 & \multicolumn{1}{c}{64} & -9.17 & \multicolumn{1}{c}{42} & \multicolumn{1}{c}{146} & -10.31 & \multicolumn{1}{c}{60} \\
27 & &&& -9.29 & \multicolumn{1}{c}{44} & \multicolumn{1}{c}{104} & -10.60 & \multicolumn{1}{c}{64} & -9.40 & \multicolumn{1}{c}{46} & \multicolumn{1}{c}{104} & -10.67 & \multicolumn{1}{c}{65} & -9.23 & \multicolumn{1}{c}{43} & \multicolumn{1}{c}{104} & -10.29 & \multicolumn{1}{c}{60} \\
30 & &&& -9.92 & \multicolumn{1}{c}{54} & \multicolumn{1}{c}{61} & -10.58 & \multicolumn{1}{c}{64} & -10.37 & \multicolumn{1}{c}{61} & \multicolumn{1}{c}{61} & -10.65 & \multicolumn{1}{c}{65} & -10.02 & \multicolumn{1}{c}{55} & \multicolumn{1}{c}{61} & -10.29 & \multicolumn{1}{c}{60} \\
33 & &&& -10.79 & \multicolumn{1}{c}{67} & \multicolumn{1}{c}{48} & -10.76 & \multicolumn{1}{c}{67} & -10.31 & \multicolumn{1}{c}{60} & \multicolumn{1}{c}{48} & -10.80 & \multicolumn{1}{c}{67} & -10.25 & \multicolumn{1}{c}{59} & \multicolumn{1}{c}{48} & -10.34 & \multicolumn{1}{c}{60} \\
	\end{tabular}
}
\end{table*}

\begin{table*}
    \caption{Results of effective magnetic field measurement from the multi-line slope method ($B_{{\rm ms}}$) and from the slope method applied on all the simulated data points in the spectral region from 500 nm to 600 nm ($B_{\rm {slope}}$) vs effective magnetic field strength. The rotational velocity is ${\rm v}$\,${\rm sin(i)}$=3\,${\rm km}$\,${\rm s^{-1}}$.}
    \label{table:dip}
    \centering
	\resizebox{\textwidth}{!}{
    \begin{tabular}{ r@{   }r@{   }r@{   }r r@{   }r@{   }r@{   }r@{   }r r@{   }r@{   }r@{   }r@{   }r r@{   }r@{   }r@{   }r@{   }r  }
        \hline\hline\\[-2.0ex]
        \multicolumn{4}{c}{ } &
        \multicolumn{5}{c}{CAOS}  &
        \multicolumn{5}{c}{NARVAL} &
       \multicolumn{5}{c}{HARPSpol}\\

       \multicolumn{4}{c}{} &
        \multicolumn{5}{c}{R=$55000$} &
        \multicolumn{5}{c}{R=$65000$} &
        \multicolumn{5}{c}{R=$115000$} \\

       \multicolumn{4}{c}{} &
       \multicolumn{5}{c}{2.5 px per FWHM} &
        \multicolumn{5}{c}{2.5 px per FWHM} &
        \multicolumn{5}{c}{4.1 px per FWHM} \\

       \multicolumn{4}{c}{} &
        \multicolumn{5}{c}{} &
       \multicolumn{5}{c}{} &
       \multicolumn{5}{c}{} \\

        \multicolumn{4}{c}{} &
        \multicolumn{5}{c}{} &
       \multicolumn{5}{c}{} &
       \multicolumn{5}{c}{} \\

       \hline\\[-1.0ex]
B$_{\rm inp}$& &&& B$_{\rm ms}$ &$\Delta$B$_{\rm inp - ms} $ & Lines & B$_{\rm slope}$                    & $\Delta$B$_{\rm inp - slope} $     & B$_{\rm ms}$  & $\Delta$B$_{\rm inp - ms} $   &Lines                & B$_{\rm slope}$            & $\Delta$B$_{\rm inp - slope} $     & B$_{\rm ms}$     & $\Delta$B$_{\rm inp - ms} $        & Lines         & B$_{\rm slope}$             & $\Delta$B$_{\rm inp - slope} $   \\
${\rm [G]}$  & &&& ${\rm [G]}$  & \multicolumn{1}{c}{(\%)}   &       & \multicolumn{1}{c}{${\rm [G]}$}     & \multicolumn{1}{c}{(\%)}         & ${\rm [G]}$   & \multicolumn{1}{c}{(\%)}    &                 & \multicolumn{1}{c}{${\rm [G]}$}  & \multicolumn{1}{c}{(\%)}         & ${\rm [G]}$      & \multicolumn{1}{c}{(\%)}         &              & \multicolumn{1}{c}{${\rm [G]}$}       & \multicolumn{1}{c}{(\%)}\\

        \hline\\[-1.0ex]
0.63 	& &&& 0.66 	& \multicolumn{1}{c}{5} 	& \multicolumn{1}{c}{1344} & 0.91 & \multicolumn{1}{c}{43}  & 0.72 & \multicolumn{1}{c}{13}  & \multicolumn{1}{c}{1344} & 0.93 & \multicolumn{1}{c}{46}  & 0.70 & \multicolumn{1}{c}{10}  & \multicolumn{1}{c}{1344} & 0.89 & \multicolumn{1}{c}{40} \\
6.35 	& &&& 7.37 	& \multicolumn{1}{c}{16} 	& \multicolumn{1}{c}{1342} & 9.23 & \multicolumn{1}{c}{45}  & 7.17 & \multicolumn{1}{c}{13}  & \multicolumn{1}{c}{1342} & 9.09 & \multicolumn{1}{c}{43}  & 7.07 & \multicolumn{1}{c}{11}  & \multicolumn{1}{c}{1342} & 8.87 & \multicolumn{1}{c}{40} \\
65 	& &&& 75 	& \multicolumn{1}{c}{18} 	& \multicolumn{1}{c}{1348} & 91   & \multicolumn{1}{c}{44}  & 75   & \multicolumn{1}{c}{18}  & \multicolumn{1}{c}{1348} & 91   & \multicolumn{1}{c}{43}  & 73   & \multicolumn{1}{c}{15}  & \multicolumn{1}{c}{1348} & 88   & \multicolumn{1}{c}{39} \\
635 	& &&& 684 	& \multicolumn{1}{c}{8} 	& \multicolumn{1}{c}{1341} & 819  & \multicolumn{1}{c}{29}  & 681  & \multicolumn{1}{c}{7}   & \multicolumn{1}{c}{1341} & 815  & \multicolumn{1}{c}{28}  & 649  & \multicolumn{1}{c}{2}   & \multicolumn{1}{c}{1341} & 788  & \multicolumn{1}{c}{24} \\
3175 	& &&& 2323 	& \multicolumn{1}{c}{-27}	& \multicolumn{1}{c}{675}  & 2740 & \multicolumn{1}{c}{-14} & 2155 & \multicolumn{1}{c}{-32} & \multicolumn{1}{c}{675}  & 2495 & \multicolumn{1}{c}{-21} & 1516 & \multicolumn{1}{c}{-52} & \multicolumn{1}{c}{675}  & 1697 & \multicolumn{1}{c}{-47} \\
        \end{tabular}
}
\end{table*}

\begin{table*}
	\caption{Results of effective magnetic field measurement from the multi-line slope method ($B_{\rm ms}$) vs S/N ratio. The input effective magnetic field is $B_{inp} = 6.35$\,G and ${\rm v}$\,${\rm sin(i)}$=3\,${\rm km}$\,${\rm s^{-1}}$. $\sigma$B$_{\rm ms}$ is the standard deviation of the measures in the simulation (details in the text). The spectral region is between 500 nm to 600 nm.}
	\label{table:sn}
	\centering 
	\begin{tabular}{ r@{   }r@{   }r@{   }r r@{   }r@{   }r@{   }r r@{   }r@{   }r@{   }r r@{   }r@{   }r@{   }r  }
		\hline\hline\\[-2.0ex]
		\multicolumn{4}{c}{ } &
		\multicolumn{4}{c}{CAOS}  &
		\multicolumn{4}{c}{NARVAL} &
		\multicolumn{4}{c}{HARPSpol}\\
		 
		\multicolumn{4}{c}{} & 
		\multicolumn{4}{c}{R=$55000$} & 
		\multicolumn{4}{c}{R=$65000$} &
		\multicolumn{4}{c}{R=$115000$} \\

		\multicolumn{4}{c}{} & 
		\multicolumn{4}{c}{2.5 px per FWHM} & 
		\multicolumn{4}{c}{2.5 px per FWHM} &
		\multicolumn{4}{c}{4.1 px per FWHM} \\
		  
		\multicolumn{4}{c}{} & 
		\multicolumn{4}{c}{} & 
		\multicolumn{4}{c}{} &
		\multicolumn{4}{c}{} \\

		\hline\\[-1.0ex]
S/N & &&& B$_{\rm ms}$ & $\sigma$B$_{\rm ms}$ & $\Delta$B$_{\rm inp - ms} $ & Lines & B$_{\rm ms}$ &  $\sigma$B$_{\rm ms}$ &  $\Delta$B$_{\rm inp - ms} $ & Lines & B$_{\rm ms}$ & $\sigma$B$_{\rm ms}$ &  $\Delta$B$_{\rm inp - ms} $ & Lines   \\
   & &&& ${\rm [G]}$  & ${\rm [G]}$          & 	\multicolumn{1}{c}{(\%)}   &       & ${\rm [G]}$  &  ${\rm [G]}$ 	  & 	\multicolumn{1}{c}{(\%)} &       & ${\rm [G]}$  & ${\rm [G]}$          & \multicolumn{1}{c}{(\%)} &    \\

		\hline\\[-1.0ex]
100 & &&& 4.18 & 5.50 & \multicolumn{1}{c}{-34} & 1274 & 6.13 & 4.88 & \multicolumn{1}{c}{-3} & 1348 & 5.98 & 1.66 & \multicolumn{1}{c}{-6} & 1537 \\ 
250 & &&& 8.13 & 1.88 & \multicolumn{1}{c}{28} & 1310 & 7.34 & 2.21 & \multicolumn{1}{c}{16} & 1375 & 6.90 & 0.63 & \multicolumn{1}{c}{9} & 1512 \\ 
500 & &&& 7.18 & 1.62 & \multicolumn{1}{c}{13} & 1322 & 7.14 & 0.66 & \multicolumn{1}{c}{12} & 1382 & 6.76 & 0.34 & \multicolumn{1}{c}{6} & 1517 \\ 
1000 & &&& 7.41 & 0.50 & \multicolumn{1}{c}{17} & 1333 & 7.13 & 0.34 & \multicolumn{1}{c}{12} & 1389 & 6.92 & 0.18 & \multicolumn{1}{c}{9} & 1519 \\ 
	\end{tabular}
\end{table*}

\begin{table*}
	\caption{Results of the effective magnetic field measurement from the multi-line slope method ($B_{{\rm ms}}$) and from the slope method applied on all the simulated data points in a region from 500 nm to 550 nm ($B_{\rm {slope}}$) in the case of decentered dipole model for the magnetic field geometry.}
	\label{table:dipcentr}
	\centering 
	\resizebox{\textwidth}{!}{
	\begin{tabular}{ r@{   }r@{   }r@{   }r r@{   }r@{   }r@{   }r@{   }r r@{   }r@{   }r@{   }r@{   }r r@{   }r@{   }r@{   }r@{   }r  }
		\hline\hline\\[-2.0ex]
		\multicolumn{4}{c}{ } &
		\multicolumn{5}{c}{CAOS}  &
		\multicolumn{5}{c}{NARVAL} &
		\multicolumn{5}{c}{HARPSpol}\\
		 
		\multicolumn{4}{c}{} & 
		\multicolumn{5}{c}{R=$55000$} & 
		\multicolumn{5}{c}{R=$65000$} &
		\multicolumn{5}{c}{R=$115000$} \\

		\multicolumn{4}{c}{} & 
		\multicolumn{5}{c}{2.5 px per FWHM} & 
		\multicolumn{5}{c}{2.5 px per FWHM} &
		\multicolumn{5}{c}{4.1 px per FWHM} \\

		\multicolumn{4}{c}{} & 
		\multicolumn{5}{c}{} & 
		\multicolumn{5}{c}{} &
		\multicolumn{5}{c}{} \\
		  
		\multicolumn{4}{c}{} & 
		\multicolumn{5}{c}{} & 
		\multicolumn{5}{c}{} &
		\multicolumn{5}{c}{} \\

		\hline\\[-1.0ex]
B$_{\rm inp}$& &&& B$_{\rm ms}$ &$\Delta$B$_{\rm inp - ms} $ & Lines & B$_{\rm slope}$ 		           & $\Delta$B$_{\rm inp - slope} $     & B$_{\rm ms}$  &$\Delta$B$_{\rm inp - ms} $ & Lines & B$_{\rm slope}$ 		    	& $\Delta$B$_{\rm inp - slope} $ & B$_{\rm ms}$ &$\Delta$B$_{\rm inp - ms} $& Lines & B$_{\rm slope}$ 			& $\Delta$B$_{\rm inp - slope} $   \\
${\rm [G]}$  & &&& ${\rm [G]}$  & \multicolumn{1}{c}{(\%)}     &     & \multicolumn{1}{c}{${\rm [G]}$}     & \multicolumn{1}{c}{(\%)}	 	& ${\rm [G]}$   &   \multicolumn{1}{c}{(\%)}    &    & \multicolumn{1}{c}{${\rm [G]}$}  & \multicolumn{1}{c}{(\%)}	 & ${\rm [G]}$  &  \multicolumn{1}{c}{(\%)}     &   & \multicolumn{1}{c}{${\rm [G]}$}   & \multicolumn{1}{c}{(\%)}\\
		\hline\\[-1.0ex]
-1.45        & &&& -1.34        & \multicolumn{1}{c}{-7}     & \multicolumn{1}{c}{350}     & -1.77                               & \multicolumn{1}{c}{22}             & -1.34         & \multicolumn{1}{c}{-8}     & 	\multicolumn{1}{c}{362}     & -1.78                            & \multicolumn{1}{c}{23}         & -1.26        & \multicolumn{1}{c}{-13}   & \multicolumn{1}{c}{384}     & -1.70                             &  \multicolumn{1}{c}{17} \\ 
	\end{tabular}
}
\end{table*}

To test the capabilities of the multi-line slope method we computed synthetic 
spectra using COSSAM (Codice per la Sintesi Spettrale nelle Atmosfere Magnetiche) 
\citep{Stift2012}. It is a fully parallelised code that solves the polarised
radiative transfer equation for a stellar atmosphere permeated by a magnetic field
under the assumption of local thermal equilibrium (LTE). The code calculates the 
emergent Stokes $IQUV$ spectrum integrated over the visible stellar disk. Details
of design decisions and implementation can be found in \citet{Stift1998a} and \citet{Stift1998b}.

All synthetic profiles were convolved with the respective FWHM of CAOS, NARVAL 
and HARPSpol and resampled to conform with the wavelength binnings. 
Simulations were performed considering a dipolar magnetic field geometry
 centered on the star, with ${\rm i=135}$\degr, ${\beta=50}$\degr, considering a zero phase. 

First, we tested the effects of the rotational velocity on the measurements; results are
shown in Table\,\ref{table:vel}. One can see that for very low values of rotational
velocity (lower than 5\,${\rm km}$\,${\rm s^{-1}}$), the results of the multi-line 
slope method differ with respect to the input by the order of 20\%. This difference
increases with rotational velocity, and for ${\rm v}\,{\rm sin(i)}$ > 
30 ${\rm km}$\,${\rm s^{-1}}$ it exceeds 50\%. For this we blame large rotational 
velocity values which affect the shape of the line profile and its derivative in
a non-negligible way.

A second simulation tests the effects of the field strength in the measure, in the case 
of low rotational velocity (${\rm v}$\,${\rm sin(i)}$ = 3\,${\rm km}$\,${\rm s^{-1}}$). 
We can see from Table\,\ref{table:dip} that the multi-line slope method gives results 
which differ by some 20\% from the input value for field strength less than 1000\,G. For
values larger than 1000\,G, the method underestimates the field with the discrepancy 
increasing with the spectral resolution. These findings may be explained by the fact 
that with higher resolution and higher field values Zeeman splitting is dominant and 
so the first derivative of Stokes\,$I$ is no longer simply related to Stokes\,$V$ 
through Eq.\,\ref{eqfond}.

We can conclude that the assumption of small velocity broadening made in Eq.\,\ref{eqfond}
is valid. Therefore, in the case of very low stellar rotational velocity -- ${\rm v}$\,${\rm sin(i)}$\,<\,5\,${\rm km}$\,${\rm s^{-1}}$ -- and low effective field
strength -- ${\rm B_{eff}<1kG}$ -- the multi-line slope method is a valid technique for measuring the effective
magnetic field. 

The simulations of Table\,\ref{table:vel} and Table\,\ref{table:dip} 
also show the advantage of the multi-line approach. Indeed we can note how the technique 
allows to retrieve results closer to the input, better than 20\% with respect to the 
slope method applied on all the simulated data points in the spectral region; this behaviour is systematic, except for 
the case of large field strengths.

The first two simulations were computed without consider the effects of the degradation due to the photon noise. A third simulation tests the capabilities of the multi-line slope method to 
retrieve the effective magnetic field with spectra of finite S/N ratio.

For each point in the synthetic spectra, we generated a random number, 
from a normal distribution with a mean zero and a standard deviation of one, and
 we divided it by the wanted value of S/N and by the square root of the synthetic
 Stokes\,$I$. This noise $\eta$ was added on the combination of Stokes profiles:
\begin{equation}
\begin{array}{c}
{\rm s_{1}=(I+V)+\eta}\\
{\rm s_{2}=(I-V)+\eta}
\end{array}
\end{equation}
that can be used to compute the noise synthetic profiles $\widetilde{\rm I}$ and $\widetilde{\rm V}$ through:
\begin{equation}
\begin{array}{c}
{\rm \widetilde{I}=\frac{s_{1}+s_{2}}{2}}\\
{\rm \widetilde{V}=\frac{s_{1}-s_{2}}{2}}
\end{array}
\end{equation}
Each measure was repeated 100 times with different random numbers, in order
 to compute the average  B$_{\rm ms}$ and the standard deviation $\sigma$B$_{\rm ms}$. 

Results of Table\,\ref{table:sn} show that, for low field values, the errors and the 
differences between input and results decrease at higher resolution. The simulations 
reveal that a minimum S/N of 250 is needed in Stokes\,$V$ in order to measure the 
effective magnetic field with an error lower than 3$\sigma$, in the case of CAOS 
resolution.

All the previous simulation were computed considering a dipolar magnetic
 field geometry, centered on the star. In order to test the impact of the different
 geometry, we performed a measure considering a more general model, using a decentered
 dipole \citep{Stift1974}. Results on Table\,\ref{table:dipcentr} shows that the
 method can be applied also in this case and, for this reason, we can conclude 
that choice of dipolar configuration do not impact the measure.
  
\section{Comparison with the Least Squares Deconvolution (LSD) technique}
\label{compareLSD}

Magnetic field measurements from high resolution spectropolarimetric data are often made
using \textit{Least Squares Deconvolution}. This method is based on the assumption that
all the spectral lines have the same profile and that they can be added linearly. The 
LSD method can extract an average Stokes profile that can be used for the measurement 
of the effective magnetic field through the first order moment of LSD Stokes\,$V$ 
\citep{Kochukhov2010}:
\begin{equation}
{\rm B_{eff} = - 7.145\,\times\,10^{6}\frac{\int v\,
Z^{v}\,dv}{\lambda_{0}\bar{g}\int Z^{I}\,dv} }
\end{equation}
where $Z^{v}$ is the Stokes\,$V$ LSD profile, $Z^{I}$ the Stokes\,$I$ LSD profile, 
$\lambda_{0}$ and $\bar{g}$ are (arbitrary) quantities adopted for the normalisation 
of weights of the Stokes\,$V$ LSD profile.

\begin{figure}
\centering
\includegraphics[trim=1.1cm 0.5cm 0.0cm
15.cm,clip=true,width=9cm]{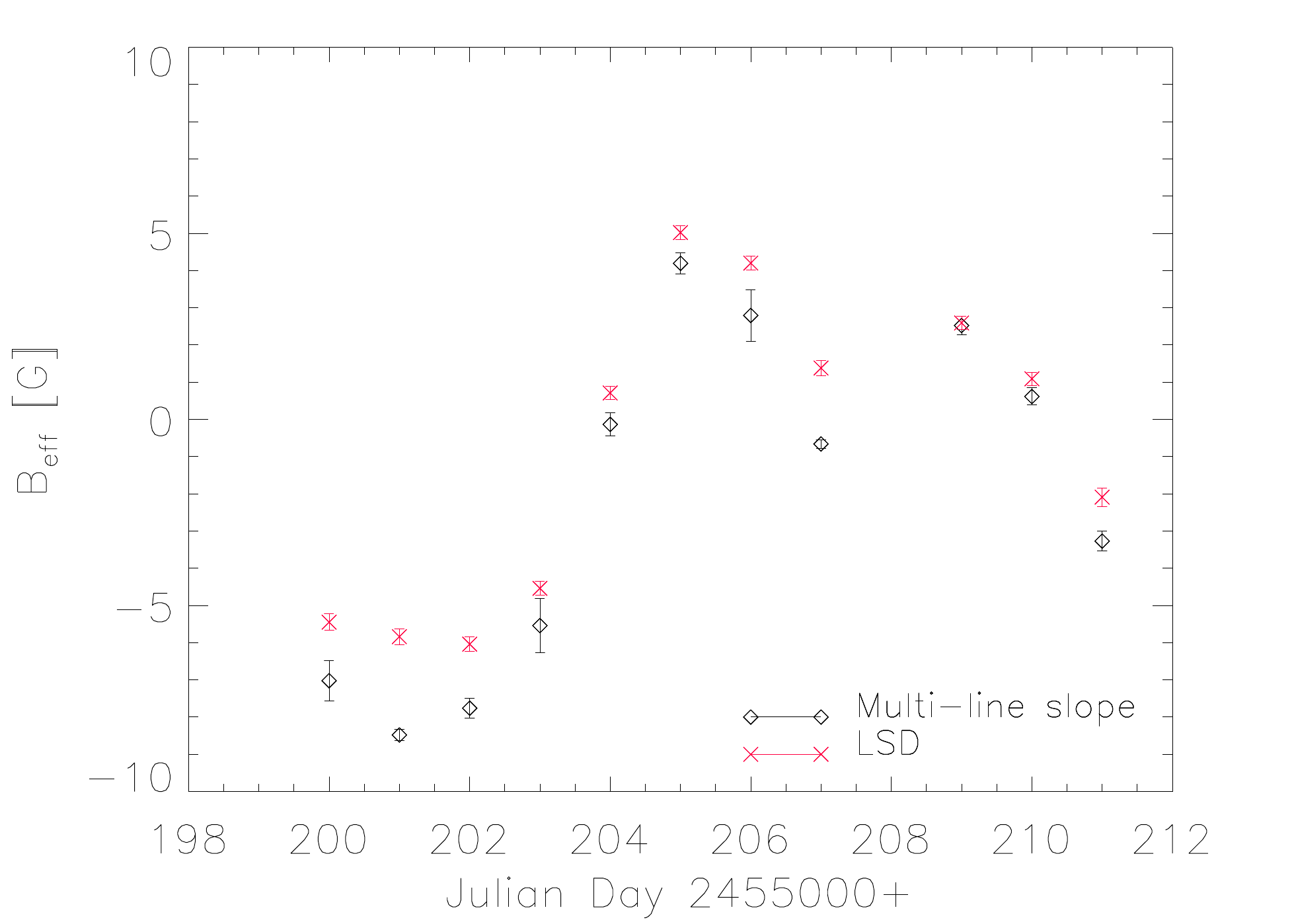}
\caption{Comparison between the multi-line slope method (black) and LSD (red) for the
$\epsilon$\,Eri HARPSpol observations of January 2010 \citep{Piskunov2011,Olander2013}.}
\label{fig:LSD_vs_slope}
\end{figure}

In order to compare the results of the multi-line slope method with the LSD results,
we measure the effective magnetic field of $\epsilon$\,Eri from the observations of
HARPSpol made in 2010, published by \citet{Piskunov2011}. The quantitative values 
of the LSD measures are taken from \citet{Olander2013}. We can see in
Fig.\,\ref{fig:LSD_vs_slope} that in general the results are in agreement; however 
the multi-line slope method gives results systematically lower than the LSD measures.
The reasons for this have yet to be determined.

We conclude that the multi-line slope method can be used as an alternative technique 
to LSD for the measurement of the effective magnetic fields of cool stars. It has the 
advantage of being easier and faster to compute; on the other hand, LSD can retrieve 
the shape of the profile, allowing the detection of magnetic field configurations with 
zero average value. At least in principle, the slope method can also
point out a zero average field when the $V/I$ scatter is larger than
$N/I$ scatter.

\section{The magnetic field of $\epsilon$\,Eridani}
\label{star}
\begin{figure*}
\centering
\includegraphics[angle=180,trim=0cm 1.0cm 3.cm
1.0cm,clip=true,width=16cm]{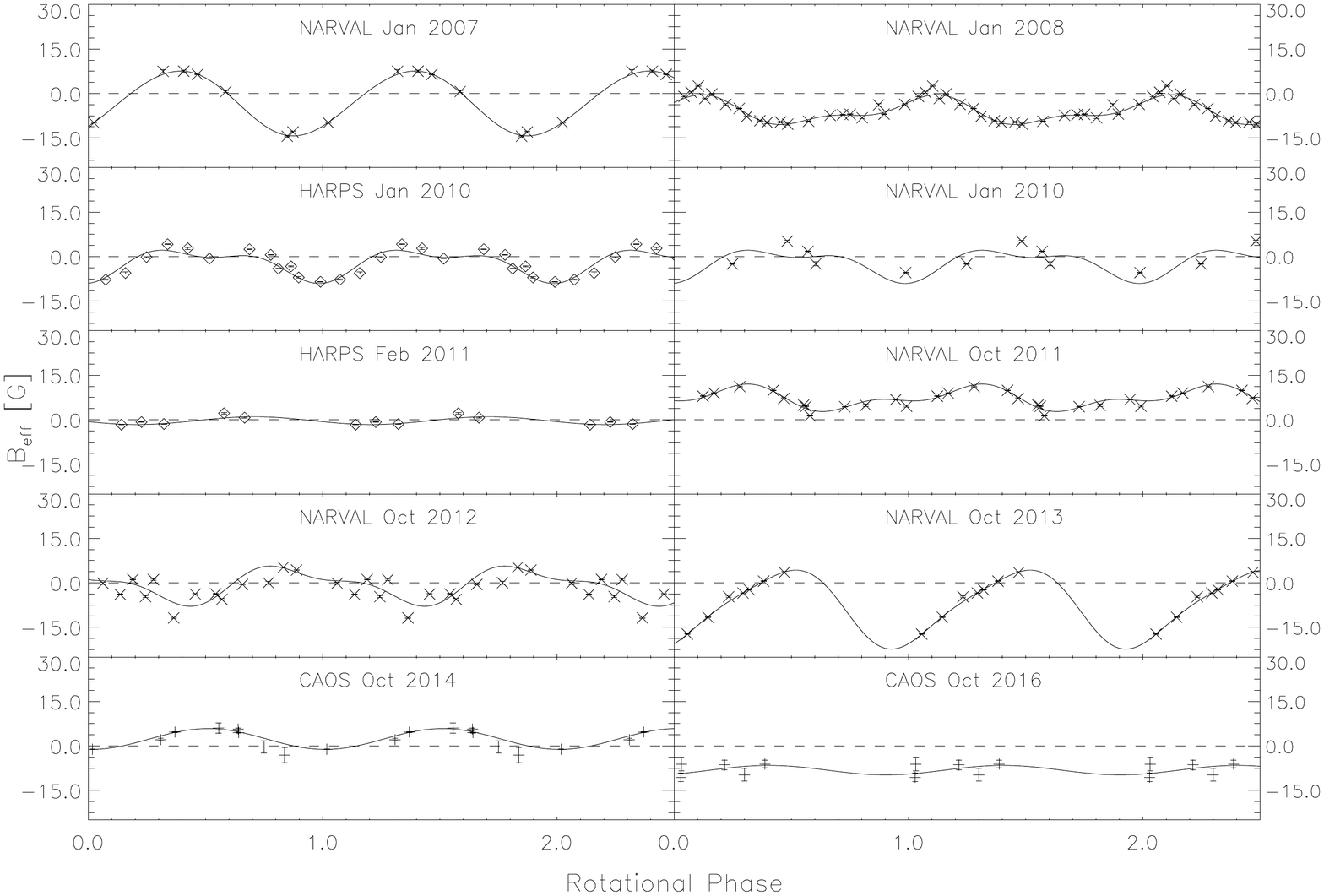}\\
\caption{Magnetic curves of $\epsilon$\,Eri folded with the period of rotation of 
the star $P_{rot}=11.35$\,d \citep{Fro2007}. Cross, diamonds and plus refers to observations obtained respectively with NARVAL, HARPSpol and CAOS. Zero phase is equal to MJD 54101 \citep{Jeffers2014}. Magnetic curves are obtained by a fitting through Eq.\,\ref{eq:azero} (see text). }
\label{fig:finalvariation}
\end{figure*}

Epsilon Eridani (HD 22049, HIP 16537) is one of the brightest and best studied solar
analogues. It is a K2V star with an effective temperature of $5146$\,K, a mass of 
0.856\,M$_{\sun}$  and ${\rm v}$ \,${\rm sin(i)} = 2.4 $\,${\rm km}$ ${\rm s^{-1}}$ 
\citep{Valenti2005}. Infrared observations show that the star is surrounded by a 
debris disk \citep{Greaves2005} with density inhomogeneities that can be explained 
by the existence of exoplanets \citep{Backman2009}.

Actually, the existence of planets in the system orbiting $\epsilon$\,Eri is an open
question. \citet{Hatzes2000} ascribed the long-period radial velocity (RV) variation 
to the presence of a planetary companion. This presence was confirmed by 
\citet{Benedict2006} who estimated a period $P_{\rm orbital} = 6.85 \pm 0.03$\,yr 
and a companion mass of $M = 1.55 \pm 0.24$\,M$_J$. \citet{Anglada2012} however 
concluded that the RV variability of $\epsilon$\,Eri is probably due to stellar 
activity cycles (not strictly periodic) rather than to the presence of a planet. 
The presence of a planet was not confirmed from the velocity measurements analysed 
by \citet{Zechmeister2013} nor from imaging either \citep{Janson2015,Mizuki2016}.

The stellar activity of $\epsilon$\,Eri over 45 years was studied by \citet{Metcalfe2013} 
who found a short cycle of 2.95\,yr modulated by a long cycle of 12.7\,yr. The magnetic 
field could play an important role in this star. In order to study the long term variation 
of it, \citet{Jeffers2014} performed spectropolarimetric observations; they collected 
Stokes\,$I$ and $V$ profiles at six epochs in a range of seven years, using the 
spectropolarimeters NARVAL and HARPSpol. They found that the large-scale magnetic
field is highly variable, with no common pattern over the years, and they concluded that more
observations were needed to investigate the magnetic activity of $\epsilon$\,Eri.

\begin{table*}
	\caption{Effective magnetic field measures of $\epsilon$Eri.} 
	\label{table:magneticmeasure} 
	\centering 
	\begin{tabular}{ c@{ }c r@{ $\pm$ }r c@{  }c c@{}c r@{ $\pm$ }r  c@{  }c c@{  }c  }
		\hline\hline\\[-2.0ex]
		\multicolumn{2}{c}{MJD} &
		\multicolumn{2}{c}{$B_{{\rm eff}}$ [G]}  &
		\multicolumn{2}{c}{Instrument} &
		\multicolumn{2}{c}{MJD} &
		\multicolumn{2}{c}{$B_{{\rm eff}}$ [G]} &
		\multicolumn{2}{c}{Instrument}\\
		& 
		\multicolumn{2}{c}{} &
		\multicolumn{2}{c}{} &
		\multicolumn{2}{c}{} &
		\multicolumn{2}{c}{} \\
		\hline\\[-2.0ex]   
		\hline\\[-2.0ex]
54122.256&& -12.97 & 0.21 && NARVAL & 55605.505 && 2.15 & 0.25 && HARPS \\ 
54127.316&& 7.49 & 0.60 && NARVAL & 55606.504 && 0.72 & 0.16 && HARPS \\ 
54128.313&& 7.53 & 0.31 && NARVAL & 55836.623 && 6.80 & 0.05 && NARVAL \\ 
54130.354&& 0.69 & 0.26 && NARVAL & 55838.638 && 7.94 & 0.10 && NARVAL \\ 
54133.326&& -14.41 & 0.11 && NARVAL & 55843.622 && 4.53 & 0.14 && NARVAL \\ 
54135.316&& -9.87 & 0.18 && NARVAL & 55845.503 && 4.39 & 0.09 && NARVAL \\ 
54140.328&& 6.43 & 0.12 && NARVAL & 55846.521 && 4.85 & 0.44 && NARVAL \\ 
54485.380&& -3.83 & 0.37 && NARVAL & 55850.517 && 9.00 & 0.12 && NARVAL \\ 
54487.305&& -1.00 & 0.06 && NARVAL & 55866.516 && 1.32 & 0.14 && NARVAL \\ 
54488.321&& -1.69 & 0.06 && NARVAL & 55874.464 && 11.18 & 0.12 && NARVAL \\ 
54489.325&& -3.68 & 0.21 && NARVAL & 55876.594 && 7.16 & 0.15 && NARVAL \\ 
54490.345&& -7.72 & 0.21 && NARVAL & 55877.552 && 4.95 & 0.13 && NARVAL \\ 
54491.328&& -9.77 & 0.33 && NARVAL & 55882.540 && 4.54 & 0.19 && NARVAL \\ 
54492.327&& -10.31 & 0.29 && NARVAL & 55887.432 && 9.88 & 0.08 && NARVAL \\ 
54493.339&& -9.36 & 0.08 && NARVAL & 56202.533 && 1.14 & 0.07 && NARVAL \\ 
54494.383&& -7.37 & 0.06 && NARVAL & 56203.533 && 1.07 & 0.23 && NARVAL \\ 
54495.352&& -7.01 & 0.08 && NARVAL & 56205.556 && -3.79 & 0.17 && NARVAL \\ 
54499.348&& 2.53 & 0.17 && NARVAL & 56206.543 && -3.74 & 0.10 && NARVAL \\ 
54501.331&& -5.01 & 0.06 && NARVAL & 56214.494 && -4.65 & 0.39 && NARVAL \\ 
54502.330&& -9.18 & 0.17 && NARVAL & 56224.617 && -3.89 & 0.13 && NARVAL \\ 
54503.330&& -9.60 & 0.14 && NARVAL & 56229.541 && -5.57 & 0.26 && NARVAL \\ 
54506.339&& -7.12 & 0.08 && NARVAL & 56230.540 && -0.56 & 0.19 && NARVAL \\ 
54507.284&& -8.19 & 0.27 && NARVAL & 56232.517 && 5.25 & 0.07 && NARVAL \\ 
54508.342&& -6.94 & 0.07 && NARVAL & 56238.554 && -11.85 & 0.14 && NARVAL \\ 
54509.347&& -3.64 & 0.20 && NARVAL & 56244.508 && 4.31 & 0.16 && NARVAL \\ 
54510.343&& 0.45 & 0.29 && NARVAL & 56246.476 && -0.15 & 0.38 && NARVAL \\ 
54511.350&& -0.11 & 0.07 && NARVAL & 56254.483 && 0.02 & 0.49 && NARVAL \\ 
54512.341&& -2.49 & 0.16 && NARVAL & 56555.500 && -3.44 & 0.05 && NARVAL \\ 
55199.609&& -4.04 & 0.04 && HARPS & 56556.500 && 0.56 & 0.06 && NARVAL \\ 
55200.593&& -7.02 & 0.27 && HARPS & 56557.500 && 3.50 & 0.23 && NARVAL \\ 
55201.650&& -8.48 & 0.08 && HARPS & 56575.500 && -17.29 & 0.10 && NARVAL \\ 
55202.593&& -7.74 & 0.13 && HARPS & 56576.500 && -11.61 & 0.14 && NARVAL \\ 
55203.550&& -5.54 & 0.37 && HARPS & 56577.500 && -4.70 & 0.28 && NARVAL \\ 
55204.569&& -0.20 & 0.16 && HARPS & 56578.500 && -2.34 & 0.13 && NARVAL \\ 
55205.593&& 4.19 & 0.14 && HARPS & 56921.608 && 6.02 & 1.64 && CAOS \\ 
55206.570&& 2.65 & 0.35 && HARPS & 56922.557 && 5.67 & 0.42 && CAOS \\ 
55207.617&& -0.66 & 0.06 && HARPS & 56922.578 && 4.52 & 0.20 && CAOS \\ 
55209.558&& 2.52 & 0.13 && HARPS & 56941.499 && 2.12 & 0.49 && CAOS \\ 
55210.587&& 0.62 & 0.11 && HARPS & 56946.504 && -0.36 & 2.12 && CAOS \\ 
55211.575&& -3.27 & 0.14 && HARPS & 56947.504 && -3.12 & 2.54 && CAOS \\ 
55224.311&& -5.36 & 0.08 && NARVAL & 56949.551 && -1.12 & 0.16 && CAOS \\ 
55231.306&& -2.42 & 0.72 && NARVAL & 57044.323 && 4.65 & 0.15 && CAOS \\ 
55241.276&& 5.21 & 0.31 && NARVAL & 57630.549 && -10.58 & 1.39 && CAOS \\ 
55242.271&& 1.78 & 0.12 && NARVAL & 57630.580 && -6.10 & 2.34 && CAOS \\ 
55600.534&& -1.66 & 0.05 && HARPS & 57634.617 && -6.12 & 1.31 && CAOS \\ 
55601.509&& -0.78 & 0.09 && HARPS & 57723.443 && -6.41 & 1.64 && CAOS \\ 
55602.599&& -1.44 & 0.04 && HARPS & 57724.413 && -9.74 & 2.13 && CAOS \\ 
	\end{tabular}
\end{table*}

The multi-line slope method measures of all available high resolution spectropolarimetric
data are reported in Table \ref{table:magneticmeasure}. Fig.\,\ref{fig:finalvariation}
shows the highly variable behaviour of the effective magnetic field of the star, folded
with the rotational period; we noted that some curves are characterised by a change of 
polarity of the field (Narval 2007, HARPSpol 2010 and Narval 2013), in others the magnetic 
field has the same sign as in the case of NARVAL 2011.
Sinusoidal fits, showed in Fig.\,\ref{fig:finalvariation}, are obtained using: 
\begin{equation}\label{eq:azero}
{\rm f(t)=A_{0}+A_{1}\, sin\left(2\pi\,\frac{t-t_{0}}{P}+A_{2}\right)+A_{3}\, sin\left(4\pi\,\frac{t-t_{0}}{P}+A_{4}\right) }
\end{equation}
where ${\rm t}$ is the time in days, ${\rm t_0}$ is a reference time 
equal to MJD 54101 \citep{Jeffers2014}, ${\rm P}$ is the variability period 
assumed as the rotational one, ${\rm A_{1}}$ and ${\rm A_{3}}$ are amplitudes
 (expressed in Gauss), ${\rm A_{2}}$ and ${\rm A_{4}}$ are phase shifts and 
${\rm A_{0}}$ represents the level of the variation of the curves (in Gauss);
 fit of the data of CAOS and HARPS\,2011 are obtained using a single wave 
(${\rm A_{3}}={\rm A_{4}}=0$) because the few number of measured points in 
their magnetic curves and the magnetic curve of Jan\,2010 is obtained using 
both HARPSpol and NARVAL data.

\begin{figure}
\centering
\includegraphics[angle=180,trim=1.6cm 4.0cm 3.4cm 0cm,clip=true,width=11cm]{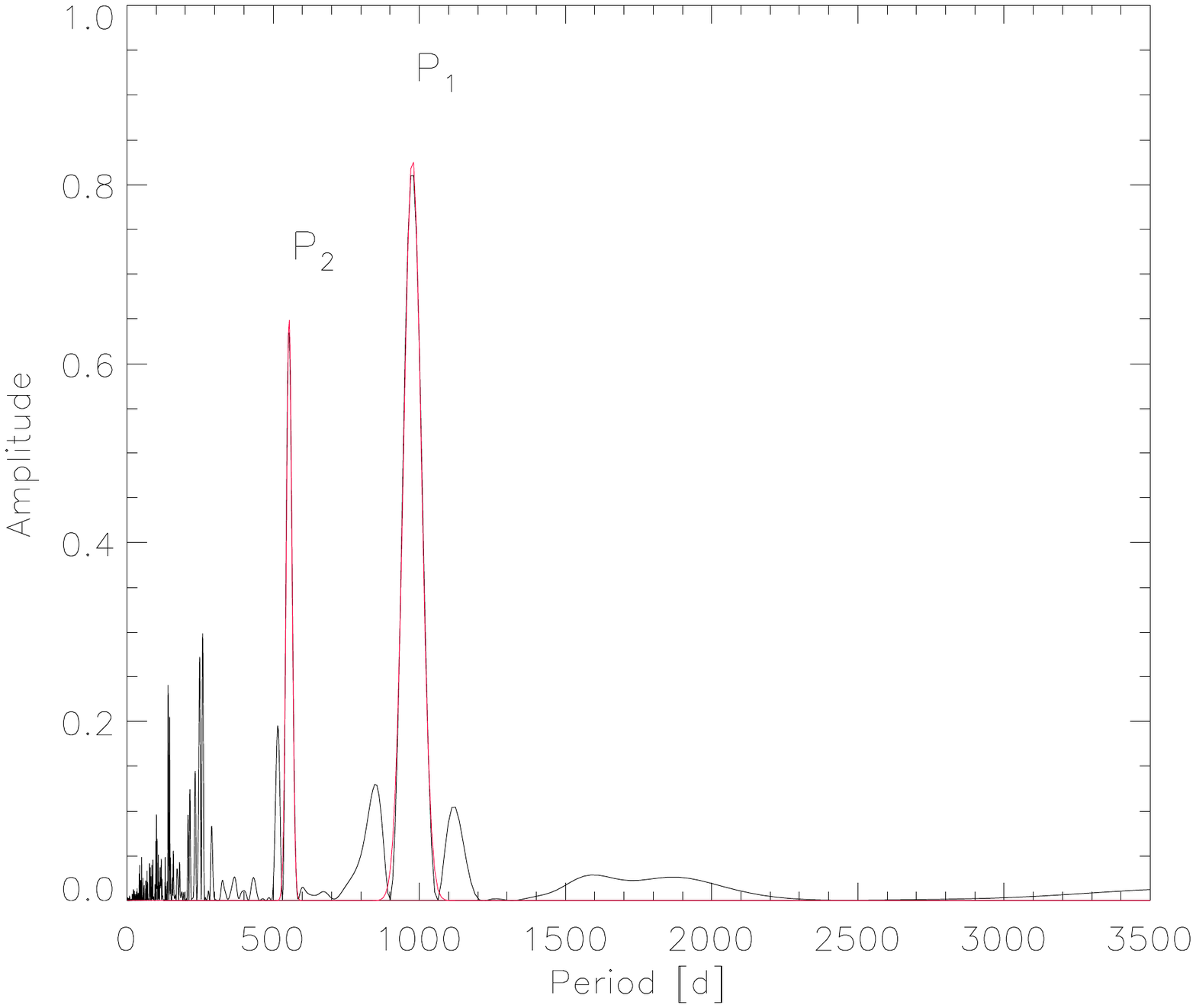}\\
\includegraphics[angle=180,trim=0.cm 1.0cm 2.cm 0.cm,clip=true,width=9.0cm]{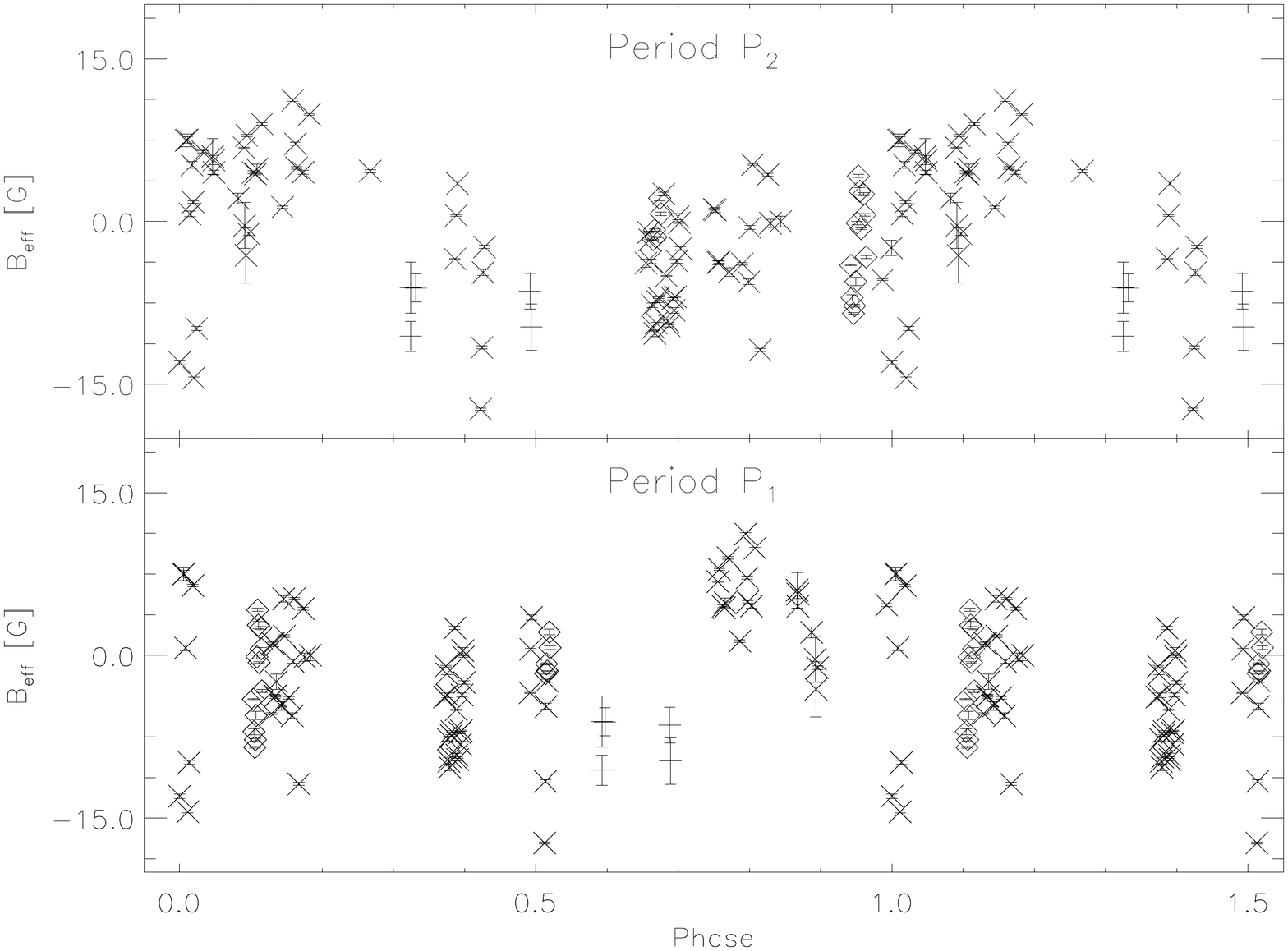}\\
\caption{{ Top}: Cleaned Fourier transform of the effective magnetic field of
$\epsilon$\,Eri (black) \citep{Deeming1975,Roberts1987} and gaussian fit of the main periods (red). {Bottom}: Effective magnetic field measures (Table \ref{table:magneticmeasure}) folded with 
the two periods. Cross, diamonds and plus refers to observations obtained respectively with NARVAL, HARPSpol and CAOS. }
\label{fig:magneticperiodigram}
\end{figure}

\begin{figure}
\centering
\includegraphics[angle=180,trim=1.5cm 3.3cm 1.9cm 1.cm,clip=true,width=10cm]{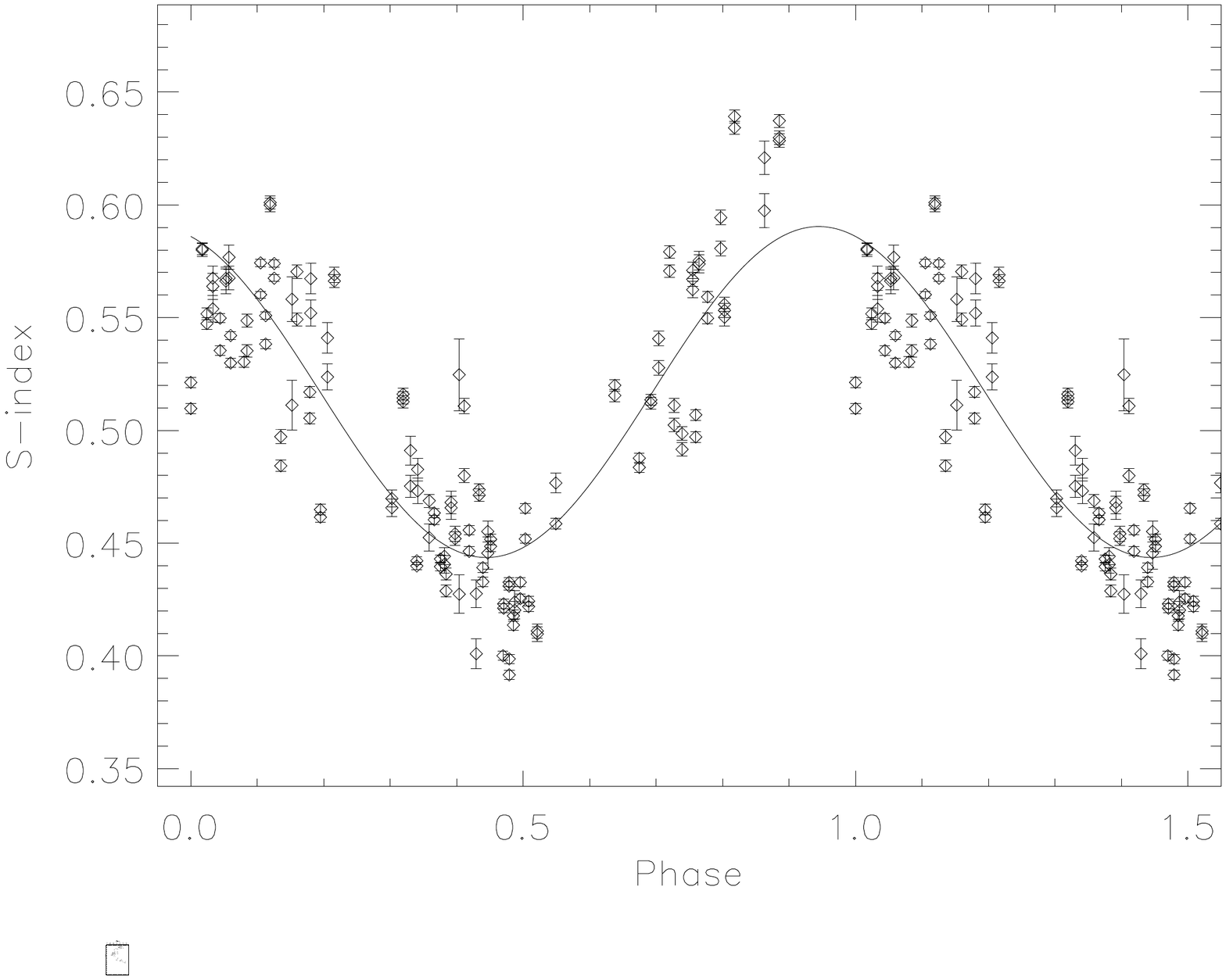}\\
\caption{S-index measures \citep{Metcalfe2013} folded with the period $P_1$. }
\label{fig:variationsindex}
\end{figure}

In order to find periodicities in the effective magnetic field we computed the 
Fourier transform following \citet{Deeming1975}. We deconvolved in the frequency domain 
using the {\sc CLEAN} algorithm \citep{Roberts1987} in order to limit the effects of 
artefacts caused by the incompleteness of the sampling. The results are reported in 
Fig.\,\ref{fig:magneticperiodigram}. Fitting a Gaussian, we estimated the position of 
the main peaks at $P_1 = 976 \pm 70$\,d and $P_2 = 555 \pm 22$\,d; errors are assumed 
to be as large as the FWHM of the Gaussian. We noted that $P_1$ is close to the 
2.95\,yr period found by \citet{Metcalfe2013} (Fig.\,\ref{fig:variationsindex}). Indeed,
 the period corresponds to that one found by \citet{Lehmann2015} analysis the magnetic field 
from the Zeeman broadening. 

\begin{table}
	\caption{Table of the values of A$_{0}$.}  
	\label{table:azerotab}
	\centering 
	\begin{tabular}{c@{ }c r@{ $\pm$ }c  }
		\hline\\[-2.0ex]
		\multicolumn{2}{c}{MJD$_{\rm average}$} & 
		\multicolumn{2}{c}{A$_{0}$} \\

		\multicolumn{2}{c}{} & 
		\multicolumn{2}{c}{[G]} \\
		\hline\\[-2.0ex]
       54131.030 && -3.15 & 0.21 \\ 
       54498.337 && -6.08 & 0.03 \\ 
       55205.339 && -2.19 & 0.04 \\ 
       55603.330 && -0.31 & 0.04 \\ 
       55860.543 && 7.12 & 0.03 \\ 
       56225.684 && -0.58 & 0.04 \\ 
       56568.214 && -8.27 & 0.67 \\ 
       56949.516 && 2.34 & 0.67 \\ 
       57668.720 && -8.17 & 0.67 \\ 
		\hline\\[-2.0ex]
\end{tabular}
\end{table}

\begin{figure}
\centering
\includegraphics[angle=180,trim=1.6cm 4.0cm 3.4cm 0cm,clip=true,width=11cm]{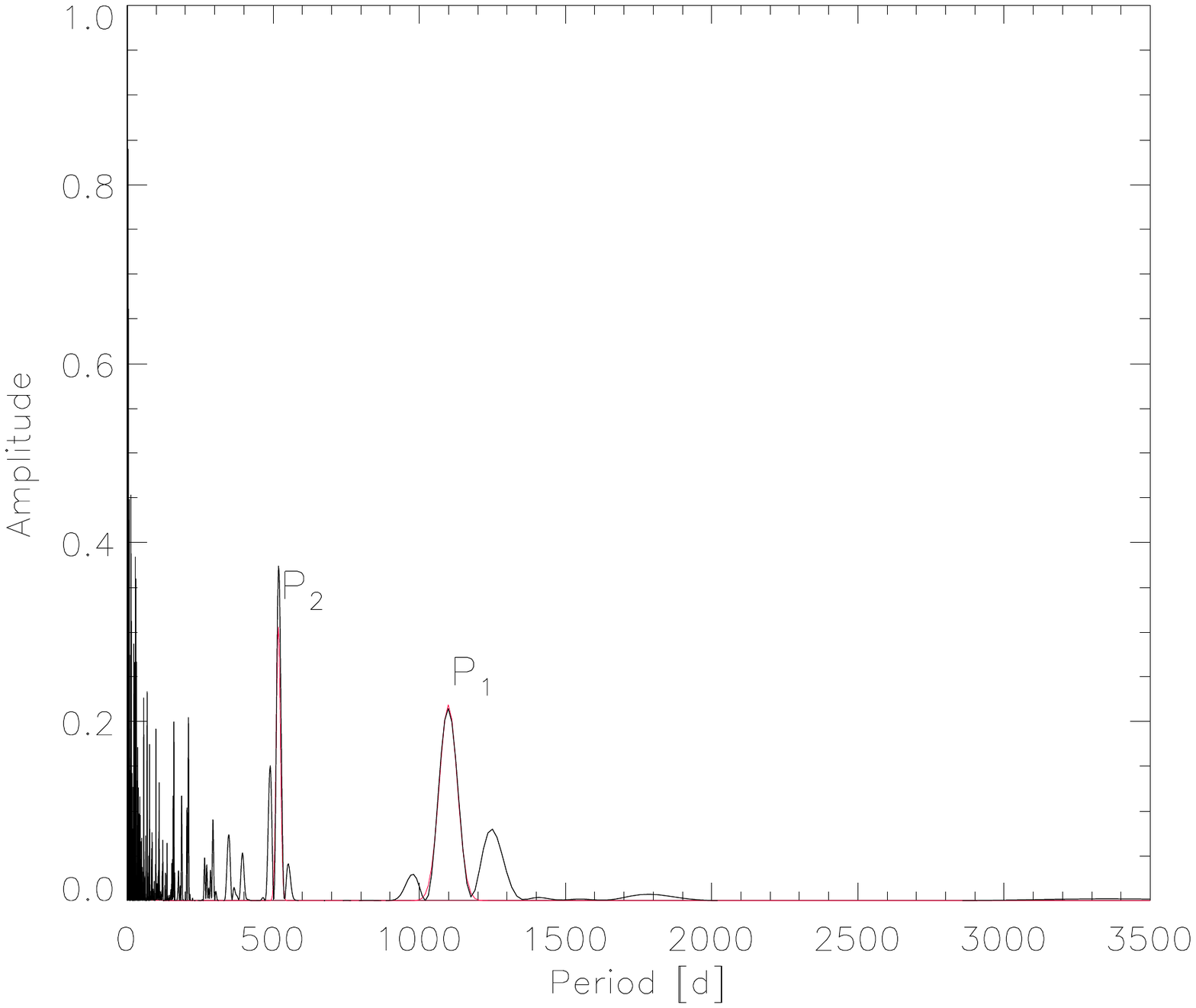}\\
\includegraphics[angle=180,trim=0.0cm 0.0cm 1.7cm
0cm,clip=true,width=8.6cm]{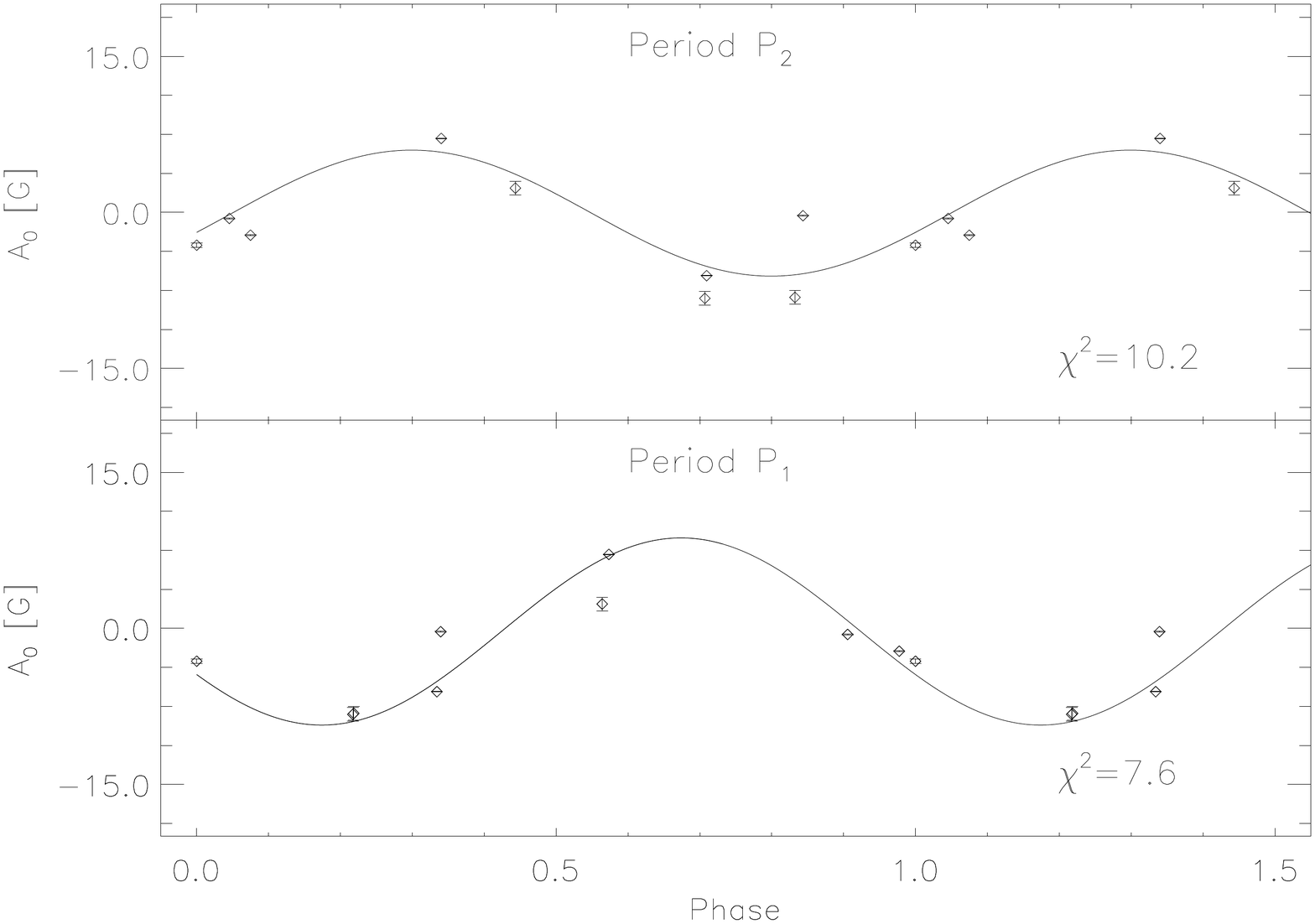}\\
\caption{{ Top}: Cleaned Fourier transform of $A_0$ (black) \citep{Deeming1975,Roberts1987} and gaussian fit of the main periods (red). {Bottom}: $A_0$ folded with the two periods.}
\label{fig:magneticperiodigram_zero}
\end{figure}

Another analysis can be performed on the variation of the level of the
 curves ${\rm A_0}$. This is done in order to separate the short-time sinusoidal variations -- due to stellar rotation -- from 
long-term changes in the constant term ${\rm A_0}$, obtained by Eq.\,\ref{eq:azero}.
The values of ${\rm A_0}$ are reported in Table\,\ref{table:azerotab} with the average time of the magnetic curves.
The Fourier transform (Fig.\,\ref{fig:magneticperiodigram_zero}) exhibits
periods near to those of the effective magnetic field at $P_1 = 1099 \pm 71$\,d 
and $P_2 = 517 \pm 17$\,d. In order to find the best period we computed $\chi^2$ 
for a sinusoidal fit to the ${\rm A_0}$ values, folded with the two periods. The results, 
reported in the bottom panel of Fig.\,\ref{fig:magneticperiodigram_zero}, reveal
that $P_1$ is the best period.

\section{Conclusion} 
\label{Conclusion}

We have introduced an extension to a method developed by \citet{Bagnulo2002} 
for measuring the effective magnetic field of stars from low resolution 
spectropolarimetry by means of a regression of Stokes\,$V$ against the
spectral derivative of Stokes\,$I$. Our \textit{multi-line slope method},
based on high resolution spectropolarimetry, instead uses similar information 
from the Stokes profiles of a large number of unblended lines. We carried 
out tests of the new method, using the polarised radiative transfer code 
{\sc Cossam}, concluding that results are satisfactory for stars with
 low rotational velocities (${\rm v}$\,${\rm sin(i)}$\,<\,5\,${\rm km}$\,${\rm s^{-1}}$)
 and field strengths (${\rm B_{eff}<1kG}$). 

The comparison with the popular Least Square Deconvolution shows that the 
multi-line slope method can be an easy-to-use and fast alternative for 
measuring the effective magnetic field of late-type stars. LSD on the other
hand offers the advantage of retrieving the shape of the Stokes profile; 
this makes it possible to infer the presence of a magnetic field with zero
average value.

Finally, we applied the technique to all the available spectropolarimetric 
data of the star $\epsilon$\,Eri. We separated the short-time variation 
of the effective magnetic field, due to stellar rotation, from the long-term 
variation through the coefficient $A_{0}$ (defined in 
Eq.\,\ref{eq:azero}); we used a Fourier transform to discover that the 
best-fit period of the variation of $A_{0}$ ($P_1 = 1099 \pm 71$ {\rm d}) is close to the short cycle 
period in the S-index found by \citet{Metcalfe2013} and to the period of the
 magnetic field found by the analysis of the Zeeman broadening \citep{Lehmann2015}. Direct measurements
of the effective magnetic field thus open up the possibility to determine 
the periods of the cycles of active cool stars.

\section*{Acknowledgements}

This research is based on observations collected at the European Organisation for 
Astronomical Research in the Southern Hemisphere under ESO programmes 084.D-0338(A) 
and 086.D-0240(A). This research has used the PolarBase database.




\bibliographystyle{mnras}
\bibliography{paper_Eps} 

\begin{thebibliography}{}
\makeatletter
\relax
\def\mn@urlcharsother{\let\do\@makeother \do\$\do\&\do\#\do\^\do\_\do\%\do\~}
\def\mn@doi{\begingroup\mn@urlcharsother \@ifnextchar [ {\mn@doi@}
  {\mn@doi@[]}}
\def\mn@doi@[#1]#2{\def\@tempa{#1}\ifx\@tempa\@empty \href
  {http://dx.doi.org/#2} {doi:#2}\else \href {http://dx.doi.org/#2} {#1}\fi
  \endgroup}
\def\mn@eprint#1#2{\mn@eprint@#1:#2::\@nil}
\def\mn@eprint@arXiv#1{\href {http://arxiv.org/abs/#1} {{\tt arXiv:#1}}}
\def\mn@eprint@dblp#1{\href {http://dblp.uni-trier.de/rec/bibtex/#1.xml}
  {dblp:#1}}
\def\mn@eprint@#1:#2:#3:#4\@nil{\def\@tempa {#1}\def\@tempb {#2}\def\@tempc
  {#3}\ifx \@tempc \@empty \let \@tempc \@tempb \let \@tempb \@tempa \fi \ifx
  \@tempb \@empty \def\@tempb {arXiv}\fi \@ifundefined
  {mn@eprint@\@tempb}{\@tempb:\@tempc}{\expandafter \expandafter \csname
  mn@eprint@\@tempb\endcsname \expandafter{\@tempc}}}

\bibitem[\protect\citeauthoryear{{Anglada-Escud{\'e}} \&
  {Butler}}{{Anglada-Escud{\'e}} \& {Butler}}{2012}]{Anglada2012}
{Anglada-Escud{\'e}} G.,  {Butler} R.~P.,  2012, \mn@doi [\apjs]
  {10.1088/0067-0049/200/2/15}, \href
  {http://adsabs.harvard.edu/abs/2012ApJS..200...15A} {200, 15}

\bibitem[\protect\citeauthoryear{{Auri{\`e}re}}{{Auri{\`e}re}}{2003}]{aur2003}
{Auri{\`e}re} M.,  2003, in {Arnaud} J.,  {Meunier} N.,  eds,  EAS Publications
  Series Vol. 9, EAS Publications Series. p.~105

\bibitem[\protect\citeauthoryear{{Backman} et~al.,}{{Backman}
  et~al.}{2009}]{Backman2009}
{Backman} D.,  et~al., 2009, \mn@doi [\apj] {10.1088/0004-637X/690/2/1522},
  \href {http://adsabs.harvard.edu/abs/2009ApJ...690.1522B} {690, 1522}

\bibitem[\protect\citeauthoryear{{Bagnulo}, {Szeifert}, {Wade}, {Landstreet}
  \& {Mathys}}{{Bagnulo} et~al.}{2002}]{Bagnulo2002}
{Bagnulo} S.,  {Szeifert} T.,  {Wade} G.~A.,  {Landstreet} J.~D.,   {Mathys}
  G.,  2002, \mn@doi [\aap] {10.1051/0004-6361:20020606}, \href
  {http://adsabs.harvard.edu/abs/2002A%26A...389..191B} {389, 191}

\bibitem[\protect\citeauthoryear{{Bagnulo}, {Landstreet}, {Mason}, {Andretta},
  {Silaj}  \& {Wade}}{{Bagnulo} et~al.}{2006}]{Bagnulo2006}
{Bagnulo} S.,  {Landstreet} J.~D.,  {Mason} E.,  {Andretta} V.,  {Silaj} J.,
  {Wade} G.~A.,  2006, \mn@doi [\aap] {10.1051/0004-6361:20054223}, \href
  {http://adsabs.harvard.edu/abs/2006A%26A...450..777B} {450, 777}

\bibitem[\protect\citeauthoryear{{Bagnulo}, {Landstreet}, {Fossati}  \&
  {Kochukhov}}{{Bagnulo} et~al.}{2012}]{Bagnulo2012}
{Bagnulo} S.,  {Landstreet} J.~D.,  {Fossati} L.,   {Kochukhov} O.,  2012,
  \mn@doi [\aap] {10.1051/0004-6361/201118098}, \href
  {http://adsabs.harvard.edu/abs/2012A%26A...538A.129B} {538, A129}

\bibitem[\protect\citeauthoryear{{Benedict} et~al.,}{{Benedict}
  et~al.}{2006}]{Benedict2006}
{Benedict} G.~F.,  et~al., 2006, \mn@doi [\aj] {10.1086/508323}, \href
  {http://adsabs.harvard.edu/abs/2006AJ....132.2206B} {132, 2206}

\bibitem[\protect\citeauthoryear{{Bouvier} et~al.,}{{Bouvier}
  et~al.}{2007}]{Bouvier2007}
{Bouvier} J.,  et~al., 2007, \mn@doi [\aap] {10.1051/0004-6361:20066021}, \href
  {http://adsabs.harvard.edu/abs/2007A%26A...463.1017B} {463, 1017}

\bibitem[\protect\citeauthoryear{{Carolo} et~al.,}{{Carolo}
  et~al.}{2014}]{Carolo2014}
{Carolo} E.,  et~al., 2014, \mn@doi [\aap] {10.1051/0004-6361/201323102}, \href
  {http://adsabs.harvard.edu/abs/2014A%26A...567A..48C} {567, A48}

\bibitem[\protect\citeauthoryear{{Deeming}}{{Deeming}}{1975}]{Deeming1975}
{Deeming} T.~J.,  1975, \mn@doi [\apss] {10.1007/BF00681947}, \href
  {http://adsabs.harvard.edu/abs/1975Ap%26SS..36..137D} {36, 137}

\bibitem[\protect\citeauthoryear{{Desidera} et~al.,}{{Desidera}
  et~al.}{2004}]{Desidera2004}
{Desidera} S.,  et~al., 2004, \mn@doi [\aap] {10.1051/0004-6361:20040155},
  \href {http://adsabs.harvard.edu/abs/2004A%26A...420L..27D} {420, L27}

\bibitem[\protect\citeauthoryear{{Donati}, {Semel}, {Carter}, {Rees}  \&
  {Collier Cameron}}{{Donati} et~al.}{1997}]{Donati1997}
{Donati} J.-F.,  {Semel} M.,  {Carter} B.~D.,  {Rees} D.~E.,   {Collier
  Cameron} A.,  1997, \mn@doi [\mnras] {10.1093/mnras/291.4.658}, \href
  {http://adsabs.harvard.edu/abs/1997MNRAS.291..658D} {291, 658}

\bibitem[\protect\citeauthoryear{{Dumusque} et~al.,}{{Dumusque}
  et~al.}{2012}]{Dumusque2012}
{Dumusque} X.,  et~al., 2012, \mn@doi [\nat] {10.1038/nature11572}, \href
  {http://adsabs.harvard.edu/abs/2012Natur.491..207D} {491, 207}

\bibitem[\protect\citeauthoryear{{Fr{\"o}hlich}}{{Fr{\"o}hlich}}{2007}]{Fro2007}
{Fr{\"o}hlich} H.-E.,  2007, \mn@doi [Astronomische Nachrichten]
  {10.1002/asna.200710876}, \href
  {http://adsabs.harvard.edu/abs/2007AN....328.1037F} {328, 1037}

\bibitem[\protect\citeauthoryear{{Greaves} et~al.,}{{Greaves}
  et~al.}{2005}]{Greaves2005}
{Greaves} J.~S.,  et~al., 2005, \mn@doi [\apjl] {10.1086/428348}, \href
  {http://adsabs.harvard.edu/abs/2005ApJ...619L.187G} {619, L187}

\bibitem[\protect\citeauthoryear{{Hatzes} et~al.,}{{Hatzes}
  et~al.}{2000}]{Hatzes2000}
{Hatzes} A.~P.,  et~al., 2000, \mn@doi [\apjl] {10.1086/317319}, \href
  {http://adsabs.harvard.edu/abs/2000ApJ...544L.145H} {544, L145}

\bibitem[\protect\citeauthoryear{{Hubrig}, {Scholz}, {Hamann}, {Sch{\"o}ller},
  {Ignace}, {Ilyin}, {Gayley}  \& {Oskinova}}{{Hubrig}
  et~al.}{2016}]{Hubrig2016}
{Hubrig} S.,  {Scholz} K.,  {Hamann} W.-R.,  {Sch{\"o}ller} M.,  {Ignace} R.,
  {Ilyin} I.,  {Gayley} K.~G.,   {Oskinova} L.~M.,  2016, \mn@doi [\mnras]
  {10.1093/mnras/stw558}, \href
  {http://adsabs.harvard.edu/abs/2016MNRAS.458.3381H} {458, 3381}

\bibitem[\protect\citeauthoryear{{Janson}, {Quanz}, {Carson}, {Thalmann},
  {Lafreni{\`e}re}  \& {Amara}}{{Janson} et~al.}{2015}]{Janson2015}
{Janson} M.,  {Quanz} S.~P.,  {Carson} J.~C.,  {Thalmann} C.,  {Lafreni{\`e}re}
  D.,   {Amara} A.,  2015, \mn@doi [\aap] {10.1051/0004-6361/201424944}, \href
  {http://adsabs.harvard.edu/abs/2015A%26A...574A.120J} {574, A120}

\bibitem[\protect\citeauthoryear{{Jeffers}, {Petit}, {Marsden}, {Morin},
  {Donati}  \& {Folsom}}{{Jeffers} et~al.}{2014}]{Jeffers2014}
{Jeffers} S.~V.,  {Petit} P.,  {Marsden} S.~C.,  {Morin} J.,  {Donati} J.-F.,
  {Folsom} C.~P.,  2014, \mn@doi [\aap] {10.1051/0004-6361/201423725}, \href
  {http://adsabs.harvard.edu/abs/2014A%26A...569A..79J} {569, A79}

\bibitem[\protect\citeauthoryear{{Judge} \& {Thompson}}{{Judge} \&
  {Thompson}}{2012}]{Judge2012}
{Judge} P.~G.,  {Thompson} M.~J.,  2012, in {Mandrini} C.~H.,  {Webb} D.~F.,
  eds,  IAU Symposium Vol. 286, Comparative Magnetic Minima: Characterizing
  Quiet Times in the Sun and Stars. pp 15--26 (\mn@eprint {arXiv} {1201.4625}),
  \mn@doi{10.1017/S1743921312004589}

\bibitem[\protect\citeauthoryear{{Kane} et~al.,}{{Kane}
  et~al.}{2016}]{Kane2016}
{Kane} S.~R.,  et~al., 2016, \mn@doi [\apjl] {10.3847/2041-8205/820/1/L5},
  \href {http://adsabs.harvard.edu/abs/2016ApJ...820L...5K} {820, L5}

\bibitem[\protect\citeauthoryear{{Kochukhov}, {Makaganiuk}  \&
  {Piskunov}}{{Kochukhov} et~al.}{2010}]{Kochukhov2010}
{Kochukhov} O.,  {Makaganiuk} V.,   {Piskunov} N.,  2010, \mn@doi [\aap]
  {10.1051/0004-6361/201015429}, \href
  {http://adsabs.harvard.edu/abs/2010A%26A...524A...5K} {524, A5}

\bibitem[\protect\citeauthoryear{{Kolenberg} \& {Bagnulo}}{{Kolenberg} \&
  {Bagnulo}}{2009}]{Kolenberg2009}
{Kolenberg} K.,  {Bagnulo} S.,  2009, \mn@doi [\aap]
  {10.1051/0004-6361/200811591}, \href
  {http://adsabs.harvard.edu/abs/2009A%26A...498..543K} {498, 543}

\bibitem[\protect\citeauthoryear{{Kurucz}}{{Kurucz}}{1993}]{Kurucz1993}
{Kurucz} R.~L.,  1993, {SYNTHE spectrum synthesis programs and line data}

\bibitem[\protect\citeauthoryear{{Landstreet}}{{Landstreet}}{1982}]{Landstreet1982}
{Landstreet} J.~D.,  1982, \mn@doi [\apj] {10.1086/160114}, \href
  {http://adsabs.harvard.edu/abs/1982ApJ...258..639L} {258, 639}

\bibitem[\protect\citeauthoryear{{Lehmann}, {K{\"u}nstler}, {Carroll}  \&
  {Strassmeier}}{{Lehmann} et~al.}{2015}]{Lehmann2015}
{Lehmann} L.~T.,  {K{\"u}nstler} A.,  {Carroll} T.~A.,   {Strassmeier} K.~G.,
  2015, \mn@doi [Astronomische Nachrichten] {10.1002/asna.201412162}, \href
  {http://adsabs.harvard.edu/abs/2015AN....336..258L} {336, 258}

\bibitem[\protect\citeauthoryear{{Leone}}{{Leone}}{2007}]{Leone2007}
{Leone} F.,  2007, \mn@doi [\mnras] {10.1111/j.1365-2966.2007.12457.x}, \href
  {http://adsabs.harvard.edu/abs/2007MNRAS.382.1690L} {382, 1690}

\bibitem[\protect\citeauthoryear{{Leone}, {Mart{\'{\i}}nez Gonz{\'a}lez},
  {Corradi}, {Privitera}  \& {Manso Sainz}}{{Leone} et~al.}{2011}]{Leone2011}
{Leone} F.,  {Mart{\'{\i}}nez Gonz{\'a}lez} M.~J.,  {Corradi} R.~L.~M.,
  {Privitera} G.,   {Manso Sainz} R.,  2011, \mn@doi [\apjl]
  {10.1088/2041-8205/731/2/L33}, \href
  {http://adsabs.harvard.edu/abs/2011ApJ...731L..33L} {731, L33}

\bibitem[\protect\citeauthoryear{{Leone} et~al.,}{{Leone}
  et~al.}{2016}]{Leone2016}
{Leone} F.,  et~al., 2016, \mn@doi [\aj] {10.3847/0004-6256/151/5/116}, \href
  {http://adsabs.harvard.edu/abs/2016AJ....151..116L} {151, 116}

\bibitem[\protect\citeauthoryear{{Metcalfe} et~al.,}{{Metcalfe}
  et~al.}{2013}]{Metcalfe2013}
{Metcalfe} T.~S.,  et~al., 2013, \mn@doi [\apjl] {10.1088/2041-8205/763/2/L26},
  \href {http://adsabs.harvard.edu/abs/2013ApJ...763L..26M} {763, L26}

\bibitem[\protect\citeauthoryear{{Mizuki} et~al.,}{{Mizuki}
  et~al.}{2016}]{Mizuki2016}
{Mizuki} T.,  et~al., 2016, \mn@doi [\aap] {10.1051/0004-6361/201628544}, \href
  {http://adsabs.harvard.edu/abs/2016A%26A...595A..79M} {595, A79}

\bibitem[\protect\citeauthoryear{{Morgenthaler} et~al.,}{{Morgenthaler}
  et~al.}{2010}]{Morgenthaler2010}
{Morgenthaler} A.,  et~al., 2010, in {Boissier} S.,  {Heydari-Malayeri} M.,
  {Samadi} R.,   {Valls-Gabaud} D.,  eds, SF2A-2010: Proceedings of the Annual
  meeting of the French Society of Astronomy and Astrophysics. p.~269
  (\mn@eprint {arXiv} {1012.0198})

\bibitem[\protect\citeauthoryear{Olander}{Olander}{2013}]{Olander2013}
Olander T.,  2013, The magnetic field of ε Eri

\bibitem[\protect\citeauthoryear{{Petit}, {Louge}, {Th{\'e}ado}, {Paletou},
  {Manset}, {Morin}, {Marsden}  \& {Jeffers}}{{Petit} et~al.}{2014}]{Petit2014}
{Petit} P.,  {Louge} T.,  {Th{\'e}ado} S.,  {Paletou} F.,  {Manset} N.,
  {Morin} J.,  {Marsden} S.~C.,   {Jeffers} S.~V.,  2014, \mn@doi [\pasp]
  {10.1086/676976}, \href {http://adsabs.harvard.edu/abs/2014PASP..126..469P}
  {126, 469}

\bibitem[\protect\citeauthoryear{{Pevtsov}, {Fisher}, {Acton}, {Longcope},
  {Johns-Krull}, {Kankelborg}  \& {Metcalf}}{{Pevtsov}
  et~al.}{2003}]{Pevtsov2003}
{Pevtsov} A.~A.,  {Fisher} G.~H.,  {Acton} L.~W.,  {Longcope} D.~W.,
  {Johns-Krull} C.~M.,  {Kankelborg} C.~C.,   {Metcalf} T.~R.,  2003, \mn@doi
  [\apj] {10.1086/378944}, \href
  {http://adsabs.harvard.edu/abs/2003ApJ...598.1387P} {598, 1387}

\bibitem[\protect\citeauthoryear{Piskunov, Kupka, Ryabchikova, Weiss  \&
  Jeffery}{Piskunov et~al.}{1995}]{Piskunov1995}
Piskunov N.~E.,  Kupka F.,  Ryabchikova T.~A.,  Weiss W.~W.,   Jeffery C.~S.,
  1995, A{\&}AS, 112, 525

\bibitem[\protect\citeauthoryear{{Piskunov} et~al.,}{{Piskunov}
  et~al.}{2011}]{Piskunov2011}
{Piskunov} N.,  et~al., 2011, The Messenger, \href
  {http://cdsads.u-strasbg.fr/abs/2011Msngr.143....7P} {143, 7}

\bibitem[\protect\citeauthoryear{{Preusse}, {Kopp}, {B{\"u}chner}  \&
  {Motschmann}}{{Preusse} et~al.}{2006}]{Preusse2006}
{Preusse} S.,  {Kopp} A.,  {B{\"u}chner} J.,   {Motschmann} U.,  2006, \mn@doi
  [\aap] {10.1051/0004-6361:20065353}, \href
  {http://adsabs.harvard.edu/abs/2006A%26A...460..317P} {460, 317}

\bibitem[\protect\citeauthoryear{{Queloz} et~al.,}{{Queloz}
  et~al.}{2001}]{Queloz2001}
{Queloz} D.,  et~al., 2001, \mn@doi [\aap] {10.1051/0004-6361:20011308}, \href
  {http://adsabs.harvard.edu/abs/2001A%26A...379..279Q} {379, 279}

\bibitem[\protect\citeauthoryear{{Reiners}}{{Reiners}}{2012}]{Reiners2012}
{Reiners} A.,  2012, \mn@doi [Living Reviews in Solar Physics]
  {10.12942/lrsp-2012-1}, \href
  {http://adsabs.harvard.edu/abs/2012LRSP....9....1R} {9}

\bibitem[\protect\citeauthoryear{{Roberts}, {Lehar}  \& {Dreher}}{{Roberts}
  et~al.}{1987}]{Roberts1987}
{Roberts} D.~H.,  {Lehar} J.,   {Dreher} J.~W.,  1987, \mn@doi [\aj]
  {10.1086/114383}, \href {http://adsabs.harvard.edu/abs/1987AJ.....93..968R}
  {93, 968}

\bibitem[\protect\citeauthoryear{{Sbordone}, {Bonifacio}, {Castelli}  \&
  {Kurucz}}{{Sbordone} et~al.}{2004}]{Sbordone2004}
{Sbordone} L.,  {Bonifacio} P.,  {Castelli} F.,   {Kurucz} R.~L.,  2004,
  Memorie della Societa Astronomica Italiana Supplementi, \href
  {http://adsabs.harvard.edu/abs/2004MSAIS...5...93S} {5, 93}

\bibitem[\protect\citeauthoryear{{Schrijver}, {Cote}, {Zwaan}  \&
  {Saar}}{{Schrijver} et~al.}{1989}]{Schrijver1989}
{Schrijver} C.~J.,  {Cote} J.,  {Zwaan} C.,   {Saar} S.~H.,  1989, \mn@doi
  [\apj] {10.1086/167168}, \href
  {http://adsabs.harvard.edu/abs/1989ApJ...337..964S} {337, 964}

\bibitem[\protect\citeauthoryear{{Semel} \& {Li}}{{Semel} \&
  {Li}}{1996}]{Semel1996}
{Semel} M.,  {Li} J.,  1996, \mn@doi [\solphys] {10.1007/BF00146653}, \href
  {http://adsabs.harvard.edu/abs/1996SoPh..164..417S} {164, 417}

\bibitem[\protect\citeauthoryear{{Semel}, {Ram{\'{\i}}rez V{\'e}lez},
  {Mart{\'{\i}}nez Gonz{\'a}lez}, {Asensio Ramos}, {Stift}, {L{\'o}pez Ariste}
  \& {Leone}}{{Semel} et~al.}{2009}]{Semel2009}
{Semel} M.,  {Ram{\'{\i}}rez V{\'e}lez} J.~C.,  {Mart{\'{\i}}nez Gonz{\'a}lez}
  M.~J.,  {Asensio Ramos} A.,  {Stift} M.~J.,  {L{\'o}pez Ariste} A.,   {Leone}
  F.,  2009, \mn@doi [\aap] {10.1051/0004-6361/200810428}, \href
  {http://adsabs.harvard.edu/abs/2009A%26A...504.1003S} {504, 1003}

\bibitem[\protect\citeauthoryear{{Sennhauser} \& {Berdyugina}}{{Sennhauser} \&
  {Berdyugina}}{2010}]{Sennhauser2010}
{Sennhauser} C.,  {Berdyugina} S.~V.,  2010, \mn@doi [\aap]
  {10.1051/0004-6361/201014971}, \href
  {http://adsabs.harvard.edu/abs/2010A%26A...522A..57S} {522, A57}

\bibitem[\protect\citeauthoryear{{Snik} et~al.,}{{Snik}
  et~al.}{2011}]{Snik2011}
{Snik} F.,  et~al., 2011, in {Kuhn} J.~R.,  {Harrington} D.~M.,  {Lin} H.,
  {Berdyugina} S.~V.,  {Trujillo-Bueno} J.,  {Keil} S.~L.,   {Rimmele} T.,
  eds,  Astronomical Society of the Pacific Conference Series Vol. 437, Solar
  Polarization 6. p.~237 (\mn@eprint {arXiv} {1010.0397})

\bibitem[\protect\citeauthoryear{{Stift}}{{Stift}}{1974}]{Stift1974}
{Stift} M.~J.,  1974, \mn@doi [\mnras] {10.1093/mnras/169.3.471}, \href
  {http://adsabs.harvard.edu/abs/1974MNRAS.169..471S} {169, 471}

\bibitem[\protect\citeauthoryear{Stift}{Stift}{1998}]{Stift1998b}
Stift M.~J.,  1998, (Astro)physical supercomputing: Ada95 as a safe, object
  oriented alternative.
Springer Berlin Heidelberg, Berlin, Heidelberg, pp 128--139,
  \mn@doi{10.1007/BFb0055000}, \url {http://dx.doi.org/10.1007/BFb0055000}

\bibitem[\protect\citeauthoryear{{Stift} \& {Dubois}}{{Stift} \&
  {Dubois}}{1998}]{Stift1998a}
{Stift} M.~J.,  {Dubois} P.~F.,  1998, \mn@doi [Computers in Physics]
  {10.1063/1.168624}, \href {http://adsabs.harvard.edu/abs/1998ComPh..12..150S}
  {12, 150}

\bibitem[\protect\citeauthoryear{{Stift}, {Leone}  \& {Cowley}}{{Stift}
  et~al.}{2012}]{Stift2012}
{Stift} M.~J.,  {Leone} F.,   {Cowley} C.~R.,  2012, \mn@doi [\mnras]
  {10.1111/j.1365-2966.2011.19933.x}, \href
  {http://adsabs.harvard.edu/abs/2012MNRAS.419.2912S} {419, 2912}

\bibitem[\protect\citeauthoryear{{Strugarek}, {Brun}, {Matt}  \&
  {R{\'e}ville}}{{Strugarek} et~al.}{2015}]{Strugarek2015}
{Strugarek} A.,  {Brun} A.~S.,  {Matt} S.~P.,   {R{\'e}ville} V.,  2015,
  \mn@doi [\apj] {10.1088/0004-637X/815/2/111}, \href
  {http://adsabs.harvard.edu/abs/2015ApJ...815..111S} {815, 111}

\bibitem[\protect\citeauthoryear{{Tinbergen} \& {Rutten}}{{Tinbergen} \&
  {Rutten}}{1992}]{Tinbergen1992}
{Tinbergen} J.,  {Rutten} R.,  1992

\bibitem[\protect\citeauthoryear{{Unno}}{{Unno}}{1956}]{Unno1956}
{Unno} W.,  1956, \pasj, \href
  {http://adsabs.harvard.edu/abs/1956PASJ....8..108U} {8, 108}

\bibitem[\protect\citeauthoryear{{Valenti} \& {Fischer}}{{Valenti} \&
  {Fischer}}{2005}]{Valenti2005}
{Valenti} J.~A.,  {Fischer} D.~A.,  2005, \mn@doi [\apjs] {10.1086/430500},
  \href {http://adsabs.harvard.edu/abs/2005ApJS..159..141V} {159, 141}

\bibitem[\protect\citeauthoryear{{Zechmeister} et~al.,}{{Zechmeister}
  et~al.}{2013}]{Zechmeister2013}
{Zechmeister} M.,  et~al., 2013, \mn@doi [\aap] {10.1051/0004-6361/201116551},
  \href {http://adsabs.harvard.edu/abs/2013A%26A...552A..78Z} {552, A78}

\makeatother
\end{thebibliography}



\bsp	
\label{lastpage}
\end{document}